# On the isothermal geometry of corrugated graphene sheets

Andrzej Trzęsowski

Institute of Fundamental Technological Research

Polish Academy of Sciences

Pawińskiego 5B, 02-106 Warsaw, Poland

e-mail adresses: atrzes@ippt.pan.pl, artrzes@gmail.com

**Abstract.** Variational geometries describing corrugated graphene sheets are proposed. The isothermal thermomechanical properties of these sheets are described by a 2-dimensional Weyl space. The equation that couples the Weyl geometry with isothermal distributions of the temperature of graphene sheets, is formulated. This material space is observed in a 3-dimensional orthogonal configurational point space as regular surfaces which are endowed with a thermal state vector field fulfilling the isothermal thermal state equation. It enables to introduce a non-topological dimensionless thermal shape parameter of non-developable graphene sheets. The properties of the congruence of lines generated by the thermal state vector field are discussed.

**1. Introduction**

Let us remind some statements concerning the physical properties of graphene monolayer which are important first of all from the thermomechanical point of view. It is known that crystals growth implies high temperatures ($\theta$) but the produced thermal fluctuations of atoms are, in the case of bulk crystals, unable to break atomic bonds of the 3D crystal structures. However, in the case of 2D crystals, these thermal fluctuations of atoms are too large and it makes impossible the creation of a stable crystalline structure. Nevertheless, it does not mean that they cannot be made artifcially.[16] For example, „one can grow a monolayer inside or top of another crystal (as an inherent part of a 3D system) and then remove the bulk at sufficiently low $\theta$ such that thermal fluctuations are unable to break atomic bonds even in macroscopic 2D crystals and they form them into a variety shapes„[16]. It is consistent with the so-called *Mermin-Wagner theorem* [33] according to which in an infinite 2D crystal, *thermal fluctuations* will destroy its long-range order. Moreover, „unlike graphite's surface, graphene is not flat but typically exhibits microscopic corrugations"[16]. The corrugations are roughly 1nm high and spread over distances of between 10 and 25 nm. These corrugations have been observed on all suspended and supported free-standing graphene sheets.([14], [28]) Note also that although „Ultraflat graphene" with ripples a few angstroms in height and several nanometers in length, much smaller than the typical sample size, can be produced, their existence can be considered as a consequence of the Mermin-Wagner theorem.[25]

Importantly, the 2D crystals (and first of all the graphene) were found not only to be continuous but to exhibit *high crystal quality*. It is perhaps because the 2D crystallites, being extracted from 3D materials, are quenched in a *metastable* state, whereas their small size and strong interatomic bonds ensure that *thermal fluctuations* cannot lead to the generation of dislocations or other crystal defects even at elevated temperature. A complementary viewpoint is that the extracted 2D crystals become *intrinsically stable* due to crumpling in the third dimension. Such 3D warping leads to a gain in *elastic energy* but suppresses thermal vibrations (anomalously large in 2D), which above a certain temperature can minimize the total free energy.([15], [28]) Note also that many distinctive electronic and chemical properties of graphene have been attributed to the presence of these ripples.[28]

We see that the existence of corrugated graphene sheets is a physical phenomenon that clearly requires to take into account the occurrence of a correlation of surface curvature not only with the internal energy of the surface but also with its thermal state.

Next, let's notice that „the most explored aspect of graphene physics is its electronic properties. It is because these are different from those of any other known condensed matter systems. First of all electrons propagating through the honeycomb graphene lattice completely lose their *effective mass*" [16]. Namely, „charge carriers in graphene are massless effective Dirac fermions and are described by the 2D analog of the quantum Dirac equation, with the *Fermi velocity* $v_F \approx 1 \times 10^6$ m/s plying the role of the speed of light and a 2D *pseudospin* describing two sublattices of the honneycomb lattice (this lattice consists of two interpenetrating triangular sublattices such that the sites of one sublattice are the centre of triangles defined by the other)"[16]. However, it ought to be stressed that the graphene relativistic behaviour arises not from required consistency with special relativity - or more specifically with Lorentz invariance - but simply from the *symmetry* of the honeycomb lattice.[14]

Although graphene is a single atomic plane of graphite, „it is neither a standard solid surface nor a standard molecule"[16]....„For example, unlike any other materials graphene *shrinks with increasing temperature θ* at all values of *θ* because membrane phonons dominate in 2D. Also, graphene exibits simultaneously high pliability (folds and pleats are commonly observed) and brittleness (it fractures like glass at high strains)".[16] Next, „graphene exhibits a breaking strength $\sim 40 \, \text{N/m}$, reaching the theoretical limit. Record values for room-temperature thermal conductivity ($\sim 5000 \, \text{W m}^{-1} \, \text{K}^{-1}$) and Young's modulus ($\sim 1.0 \, \text{TPa}$) were also reported. Graphene can be stretched elastically by as much as 20%, more than other crystal"[16]...."Speaking of non-electronic properties, we do not even know such basis things about graphene as how it melts. Neither melting temperature nor the order of phase transition is known. Ultrathin films are known to exhibit melting temperatures that rapidly decrease with decreasing thickness. The thermodynamics of 2D crystals in 3D space could be very different from that of thin films and may be closely resemble the physics of soft membranes. For example, melting can occur through generation of defect pairs and be dependent on lateral size".[16]

It is observed that „the shape of a free-standing graphene often tends to be crumpled or form nanoscrols, in close relation with its bending properties. Therefore, understanding the bending properties of graphene is very important for both practical and scientific points of view. It has been shown that the bending rigidity depends on the *temperature, size, edge shape* (e.g. bending stiffness of armchair graphene is different from that of zigzag graphene; see remarks below eq.(3)) and other factors....Note that while the bending modulus of graphene exhibits a slight nonlinearity as the bending curvature increases, the bending stiffness of graphene is dependent on the bending curvature".[31] It is reported also that the ripple structure can be controlled by *thermal treatment*. Moreover, graphene rippling by thermal treatment is related to its *negative thermal expansion coefficient* and its membrane-nature mechanical properties.[51]

Membranes are usually considered as 2D materials with zero bending stiffness (graphene exhibits its finite value). They can only sustain tensile loads; their inability to sustain compressive loads leads to wrinkling. Consequently, the membrane model of graphene can be useful e.g. in the case of estimation of an axial tensile-strain curve. However, if a critical buckling strain is estimated, then the model graphene as a thin plate is more appropriate.[53] In this case it ought to be taken into account that due to the inextensible but bendable nature of the graphene's bonds, its *effective mechanical thickness*

$$h_{\text{eff}} = \left(12k / E_{2D}\right)^{1/2},$$
$$k \approx 1\,\text{eV}, \quad E_{2D} \approx 2.12 \times 10^3 \,\text{eV/nm}^2, \tag{1}$$

where $k$ is the bending rigidity and $E_{2D}$ is the tensile rigidity, is less than 1Å. It is, according to this estimation, the smallest mechanical thickness ever achieved for any material.[53] The critical compressive strain, $\varepsilon_{cr}$, for the buckling of a rectangular thin shell under the uniaxial compression can be estimated by:

$$\varepsilon_{cr} \approx \frac{\pi^2}{3(1-v^2)}\left(\frac{h_{eff}}{l}\right)^2 \tag{2}$$

where $l$ is the length along which the uniaxial compression is applied and $v$ is the Poisson ratio. The above estimation is mainly valid for suspended thin films.**[13, 57]** It ought to be stressed that the above estimations have only qualitative meaning.

Thus, it seems physically reasonable to consider graphene as a *solid surface* treated as a „membrane" endowed with such *constitutive relations* that define its internal forces in a manner strongly dependent on the geometry of the considered sample and allowing the existence of its finite tensile as well as bending rigidity. Theoretically, the elastic properties of graphene can be studied in the *local continuum mechanics* approach (see e.g. [2] and [49]) while the estimations of effective mechanical properties needs also *atomistic simulations* (see e.g. [30]). The local continuum mechanics approach is applicable if the samples are macro. Graphene samples in the mechanical experiment have usually a radius $r \geq 0.75\,\mu\text{m}$.[24] For a *planar and circular* graphene sample of the radius $r = 0.75\,\mu\text{m}$, the ratio of the number $N_b$ of the boundary atoms to the number $N_i$ of the inner atoms can be estimated as [24]:

$$N_b / N_i = 1.5\sqrt{3}b / r \approx 5 \times 10^{-4}, \tag{3}$$

where *b* = 1.42 Å is the C-C bond length in the graphene. „It shows that the number of boundary atoms in the sample is about four orders lesser than the inner atoms. As a result, the contribution of the boundary atoms to the total energy is also four order smaller than the inner atoms. In this sense, we can ignore the contribution of the boundary atoms to the total energy in the graphene."[24] However, the graphene samples can be scaled to extremely fine scales, possibly down to a single benzene ring, because unlike other materials graphene remains stable and conductive at the molecular scale.[14] In this case, if we are dealing with the continuized models of graphene nanoclusters (e.g. graphene nanodisks - a nanometer-scale disk-like graphene cluster [10]), then the above approximation as well as the local continuum mechanics approach fail and models taking into account *size* and *shape effects* ought to be applied (see e.g. [47] and [48]). For example, „the *effects of edges* of graphene nanostructures can be modeled using one-dimensional periodic strips of grahene. Such models are commonly referred to as *graphene nanoribbons* (e.g. [10]). There are two high-symmetry crystallographic directions in graphene, *armchair* and *zigzag*. Cutting graphene nanoribbons along these directions produces armchair and zigzag (finite-length) nanoribbons, respectively."[54] It is predicted, that the *armchair* nanoribbons would reveal either metallic or semiconducting behavior and the two situations alternate as nanoribbon's width increases. Within the same model all *zigzag* graphene nanoribbons are metallic.[54] Notice also that the melting temperature of *small carbon flakes* is lower than those for graphene and graphene nanoribbons and the melting temperature of small flakes on average increases versus the number of atoms in these nanoclusters.[40]

Conventional three-dimensional crystal lattices are terminated by surfaces, which can demonstrate complex substructures, localized strain and dislocation formation. Two-dimensional crystal lattices are terminated by lines. The additional available dimension of such interfaces opens up a range of new topological distortions of crystal structures which are not available at „classic" surfaces of three-dimensional materials. The result is a rich variety of *potential* interface types in graphene, all of which can radically alter the properties of the material, even at quite long range from the interface.[23] For example, topological defects are formed by replacing a hexagon of the graphene sheet crystalline structure by a n-sided polygon. Particularly, it is known that a *pentagon* built between hexagons cause positive curvature while a *heptagon* cause negative curvature.[58] However, while a pentagon and a heptagon at short distances can be seen as a *dislocation* of the lattice [55] it is not the case of the so called *Stone-Wales defect*. Namely, in this topological defect four hexagons of planar graphene structure are changed into two pentagons and two heptagons, and the outcome arrangement of this structure remains planar. Several examples of the structures containing more pentagon-heptagon pairs but preserving flatness of the graphene sheet are known.[58] Notice also that a model of *amorphous graphene* can be generated by introducing Stone-Wales defects into perfect honeycomb lattice. Namely, it was realized in a *graphene nanocluster* „containing 800 atoms, each of them three-coordinated, similar to the honeycomb lattice but topologically distinct, with 34.5% of the elementary rings being pentagons, 38% hexagons, 24% heptagons and 4.5% octagons. Since the average size of rings is six, according to Euler's theorem, such a system can exists as a flat 2D structure with some distortions of bond lengths and angles".[26]

Let us quote yet interesting statements formulated in the paper [58] and concerning the lattice defects.„The *mitosis* is a lattice defect where two pentagons originate from a given hexagon and consequently the neighbouring hexagons become heptagons. The heptagons are separated by pentagons. The number of atoms increases by two. In this case the distorted graphene structure is a planar structure if the heptagon pair separated by the pentagon pair is studied alone, but the structure is not planar if this defect is constrained in the graphene structure. If the mitoses are arranged next to other along a line, the structure distortions are summarized along the line, the sum is very large, and the solutions is a *wavy pattern* with alternating curvatures. Mitosis can be arranged next to each other not only along straight lines but along curves or groups. For example: three pentagons placed next to each others produces a larger curvature in the graphene structure than in the case two pentagons. It is interesting that the largest curvature arises from six mitoses arranged in a group. In this case, six pentagons placed next to each other created a half dodecahedron, which can be the end of an armchair-type nanotube. If the six pentagons are arranged along a curve, the resulting structure is the end of a zigzag-type nanotube. The arrange more than six pentagons next to each other cannot be solved in a pentagon-heptagon-hexagon system. So, the pentagons and/or heptagons can occur in the system alone. When a pentagon is surrounded by hexagons, a *spherical surface* forms and when a heptago*n* is surrounded by hexagons, the characteristic *saddle-shaped surface* forms."

The above mentioned types of disorder have to be, in general, distinguished from dislocations and grain boundaries, structural defects characterized by the finite values of their respective topological invariants, Burgers vectors and misorientation angles.[55] In this paper we will restrict ourselves to the case when a graphene sheet is corrugated but the *internal geometry* of its continuized hexagonal-polygonal crystalline structure can be described by a *curvature tensor* only. We are leaving the analysis of the influence of another topological invariants of the graphene structural defects, for example of these appearing due to the applied technology of the production of graphene sheets, to a separate study; see, for example, [7, 31, 43]. It means, among others, that we are dealing with a *continuum model* of the bending properties of corrugated graphene sheets.

The *bending properties* of graphene sheets are critical in attaining the structural stability and morphology for both suspended and supported of these sheets, and directly affect their electronic properties.[3] Moreover, the bending properties not only control the morphology and electronic properties of (two-dimensional) graphene, but also *interplay* with its magnetic and thermal properties.[52] The Gaussian *bending stiffness* (associated with the Gaussian curvature of elastic graphene sheets) and the *bending rigidity* (associated with the mean curvature of elastic graphene sheets) are the two key parameters that govern the *rippling* of suspended graphene - an unavoidable phenomenon of two-dimensional materials when subject to a thermal or mechanical field. The determination these two parameters is of significance for both the design and the manipulation of graphene morphology for engineering applications. [52] The aim of this paper is to formulate the such geometrical model of this *thermomechanical phenomenon* that takes into account some *distortions* of the crystal structure of graphene sheets.

To study the *stability* of the *distorted structures*, the cohesive energy (the average energy of the chemical bonds: the total energy divided by the number of the bonds) was calculated in [58] for the above discussed structures. It was concluded that: „Cohesive energy in the environment of the defects increases several percent compared to the cohesive of the perfect graphene in every case. The increase is smaller for the planar structures and it is larger for the structures with curvatures. The more pentagons are connected to each other, the larger the decrease of stability. The mitoses arranged along a straight line have the least stability because the pentagons cause curvatures in both sides of the graphene sheet where they are connected with each other."[58]

Notice also that [7]: „In the context of *electronic systems* like graphene, the dynamics of the lattice defects occurs at a much higher energy than the electronic processes (in graphene, lattice processes are related to the sigma bonds and have typical energies of the order of tens of eV while the continuum relevant low energy processes are of few tens of meV). It then makes perfect sense to consider the motion of electrons in a *frozen geometry*."

The paper is organized as follows. In Section 2 is introduced the notion of a *material space* of graphene sheets. This material space describes isothermal corrugations of graphene sheets and is defined as a two-dimensional Weyl space with the geometry of variational type. The equation that couples this Weyl material geometry with isothermal distributions of the graphene temperature is formulated. In Section 3 three-dimensional orthogonal *configuration point spaces* of mechanical („spatial") or electronic („internal") configurations of graphene sheets, are considered. Moreover, in Section 3 the case of *developable graphene sheets* is discussed. The Weyl material space can be observed in the orthogonal configurational point space as developable regular surfaces or as non-developable regular surfaces with the geometry of variational type (Sections 3 and 4). These configurations of the graphene sheet are additionally endowed with a distinguished tangent *thermal state vector field* (Sections 2, 4 and 5) fulfilling the induced *isothermal thermal state equation* (Section 4). It enables, among others, to define a *non-topological dimensionless thermal shape parameter* of non-developable graphene sheets (Section 4). In Section 5 the geometry of graphene sheets (developable as well as non-developable) is analyzed in terms of the congruence of lines generated by the thermal state vector field. Particularly, a representation of the thermal shape parameter formulated in terms of the geometry of this congruence is given and the case of *flat thermal state vector field* is discussed. In Section 6 the conclusions and remarks concerning the shape and curvature ef-fects, the thermal state of graphene, and the membrane model of graphene sheets, are formulated.

**2. Weyl material space of graphene sheets**

If the existence of corrugated configurations of a graphene sheet is treated as the material property of graphene (see Section 1), then it seems physically reasonable to consider its 2D „material" geometry independent from the geometry of configuration spaces of this graphene sheet (see Section 3). So, let $M$ be a 2-dimensional manifold endowed with a metric **a** and a covariant derivative $\nabla$ and let in a general coordinate system $(U, u)$, $u = (u^1, u^2): U \subset M \to \mathrm{R}^2$ on $M$:

$$\mathbf{a} \doteq (a_{\alpha\beta}), \quad \mathbf{\Gamma} \doteq (\Gamma^{\kappa}_{\alpha\beta}), \tag{4}$$

where $\Gamma^{\kappa}_{\alpha\beta}$ are connection coefficients of the covariant derivative $\nabla$. Let as consider the variation of an appropriate action with respect to both the metric components and the connection coefficients without imposing from the beginning that $\Gamma^{\kappa}_{\alpha\beta}$ be the usual Christoffel symbols. This variational principle, where the metric and the connection are considered as independent variables, is called the Pallatini variation. Many distinctive electronic, chemical and bending properties of graphene have been attributed, as it was mentioned in Section 1, to the presence of ripples, which are also predicted to give rise to physical phenomena that would be absent in a planar two-dimensional materials. Consequently, the extracted 2D crystals can be *intrinsically stable* due to the *curvature effects* independently from the existence of topological defects of their crystalline structure (see Section 1). So, let us consider an action of the form [11]:

$$\mathbb{S}(\mathbf{\Gamma}, \mathbf{a}) = \int_M L(\mathrm{R})\sqrt{a}\,\mathrm{d}^2 u,$$
$$a = \det(a_{\alpha\beta}) = a_{11}a_{22} - a_{12}^2, \tag{5}$$

where the connection is *torsionless* (see Section 1 and e.g. [7]) and

$$R_{\alpha\beta}(\mathbf{\Gamma}) = R_{\kappa\alpha\beta}{}^{\kappa}(\mathbf{\Gamma}) \tag{6}$$

is the *Ricci tensor* of the covariant derivative $\nabla$ (in the designations of e.g. [49]), and

$$R = a^{\alpha\beta}R_{\alpha\beta}(\mathbf{\Gamma}), \quad \alpha, \beta = 1, 2, \tag{7}$$

is the *scalar curvature* of $\nabla$ and **a.** The Lagrangian $L$ has the form $L = L(\mathrm{R}) = -E(\mathrm{R})$, where the *material internal energy E* due to the *curvature effect* is a given function of one real variable which is assumed to be *analytic* on its domain of definition. This Lagrangian is proposed in the paper as a basis to the formulation of an example of the geometry which describe the *material properties* of a corrugated graphene sheet in a manner independent from its observation in the configurational space.

The *Euler-Lagrangian equations* for the action (5) with respect to independent variations of **a** and **Γ** can be written in the following form [11]:

$$L'(R) R_{(\alpha\beta)}(\Gamma) - \frac{1}{2} L(R) a_{\alpha\beta} = 0 \tag{8}$$

and

$$\nabla_\alpha \left( L'(R) \sqrt{a}\, a^{\beta\kappa} \right) = 0 , \tag{9}$$

where $\nabla_\alpha$ is the covariant derivative with respect to **Γ** and $R_{(\alpha\beta)}$ is the symmetric part of $R_{\alpha\beta}$. Taking the trace of eq.(8) one obtains the following equation for *R*:

$$RL'(R) - L(R) = 0 . \tag{10}$$

Let us assume that eq.(10) is not identically satisfied and has at least one real solution. Then, since analytic functions can have at most a discrete set of zeroes on the real line, eq.(10) has no more than a countable set of solutions $R = r_i$, $i = 1,2..$, where $r_i$ are constant. If at the point $R = r_i$ we have [11]

$$L'(r_i) \neq 0, \tag{11}$$

then eq.(9) takes the form

$$\nabla_\alpha \left( \sqrt{a}\, a^{\beta\kappa} \right) = 0, \tag{12}$$

while eq.(8) reduces to

$$R_{(\alpha\beta)}(\Gamma) = \frac{1}{2} r_i a_{\alpha\beta} . \tag{13}$$

It can be shown that eq.(12) has the Weyl connection as its general solution [11]:

$$\Gamma^\sigma_{\alpha\beta} = \Gamma^\sigma_{\alpha\beta}(\Gamma, \mathbf{w}) = \Gamma^\sigma_{\alpha\beta}(\mathbf{a}) - \frac{1}{2}\left( w_\alpha \delta^\sigma_\beta + w_\beta \delta^\sigma_\alpha - w^\sigma a_{\alpha\beta} \right) , \tag{14}$$

where

$$\Gamma^\sigma_{\alpha\beta}(\mathbf{a}) = \frac{1}{2} a^{\sigma\omega} \left( \partial_\alpha a_{\beta\omega} + \partial_\beta a_{\alpha\omega} - \partial_\omega a_{\alpha\beta} \right) \tag{15}$$

is the Levi-Civita connection for the metric **a** and $\mathbf{w} = w^\alpha \partial_\alpha \in W(M)$, where W(*M*) is the linear module of vector fields on *M* tangent to *M* (see [45] - Appendix). Then

$$\nabla^{a,w}_\sigma a_{\alpha\beta} = w_\sigma a_{\alpha\beta}, \qquad w_\sigma = a_{\sigma\omega} w^\omega , \tag{16}$$

for the Weyl covariant derivative $\nabla^{a,w}_\sigma$ defined by the connection of eq.(14) and, using eqs.(6), (7) and (13) - (15), one can see that eq.(8) reduces to the following condition [11]:

$$a^{\alpha\beta} R_{\alpha\beta}(\Gamma^{a,w}) = 2K + \nabla^a_\alpha w^\alpha = r_i, \qquad K = R(\mathbf{a})/2, \tag{17}$$

9where *K* and *R*(**a**) denotes the *Gauss curvature* and the scalar curvature of the Riemannian manifold $M_a = (M, \mathbf{a})$, respectively.

Although graphene is a single atomic plane of graphite, it is neither a standard solid surface nor a standard molecule (Section 1). For example, as it was mentioned in Section 1, unlike any other materials graphene shrinks with the increasing temperature $\theta$ at all values of $\theta$. Moreover, graphene rippling by thermal treatment is related to its negative thermal expansion coefficient and the membrane nature of mechanical properties (Section 1). Since the graphene can be assumed to be *thermally isotropic*, the thermal extension can be assumed to be the same in all directions in the any point $\mathrm{P} \in M$ and the change in the measure of length of a vector tangent to the curved graphene can be described in the framework of the *Weyl geometry*. Namely, let us denote, for a smooth vector field $\mathbf{v} \in \mathrm{W}(M)$ on $M$ tangent to $M$, the length of a vector $\mathbf{v}_\mathrm{P} = \mathbf{v}(\mathrm{P}) \in \mathrm{T}_\mathrm{P} M$ by [45]

$$l_{a,\mathrm{P}}(\mathbf{v}) = l_a(\mathbf{v}_\mathrm{P}) = l_a(\mathbf{v})(\mathrm{P}),$$
$$l_{a,\mathrm{P}(u)}^2(\mathbf{v}) = a_{\alpha\beta}(u) v^\alpha(u) v^\beta(u), \qquad \mathbf{v}_{\mathrm{P}(u)} = v^\alpha(u) \partial_{\alpha|\mathrm{P}(u)}, \tag{18}$$

where $\mathrm{P} = \mathrm{P}(u) \in M$ iff $u = u(\mathrm{P}) \in \mathbb{R}^2$ in terms of the coordinate description $(U, u)$ of the manifold $M$. Next, let δ denotes the *infinitesimal variation operator* defined by the Weyl covariant derivative (see e.g. [45] - Appendix):

$$\delta = \mathrm{d}u^\alpha \nabla_\alpha^{a,w}. \tag{19}$$

Then, according to eq.(16), we have:

$$\delta l_a^2(\mathbf{v}) = 2 l_a(\mathbf{v}) \mathrm{d}l_a(\mathbf{v}), \qquad \mathbf{v} = v^\alpha \partial_\alpha \in \mathrm{W}(M),$$
$$= 2 v_\alpha \delta v^\alpha + \delta a_{\alpha\beta} v^\alpha v^\beta, \qquad \delta a_{\alpha\beta} = a_{\alpha\beta} \mathrm{w}. \tag{20}$$

If $\mathbf{v} \in \mathrm{W}(M)$ is a *covariant constant*, that is

$$\delta \mathbf{v} = (\delta v^\alpha) \partial_\alpha = 0, \qquad \delta v^\alpha = \mathrm{d}u^\kappa \nabla_\kappa^{a,d} v^\alpha, \tag{21}$$

then, independently of the choice of this vector field, we have

$$\frac{1}{2} \frac{\delta l_a^2(\mathbf{v})}{l_a^2(\mathbf{v})} = \frac{\mathrm{d}l_a(\mathbf{v})}{l_a(\mathbf{v})} = \varepsilon, \qquad \varepsilon = \frac{1}{2} \mathrm{w},$$
$$\mathrm{w} = w_\alpha \mathrm{d}u^\alpha \in \mathrm{W}(U)^*, \qquad w_\alpha = a_{\alpha\beta} w^\beta, \tag{22}$$

Thus, along a curve $\gamma : I \to M$ such that for its tangent vector field $\dot\gamma$ and for a covariant constant tangent vector field $\mathbf{v} \in \mathrm{W}(M)$, we have

$$\dot\gamma = \mathbf{v} \circ \gamma : t \in I \mapsto \dot\gamma(t) = \mathbf{v}(\gamma(t)) \in \mathrm{T}_{\gamma(t)} M, \qquad \delta \mathbf{v} = 0, \tag{23}$$

the following relation holds:

$$l_a = l_0 \exp\left(\int_\gamma \varepsilon\right), \qquad [l_0] = \text{cm} . \tag{24}$$

It follows that the *length measurement* in the Weyl space depends on the way along this measurement is done.

If the intrinsic material metric tensor **a** undergoes a *conformal transformation*, i.e.,

$$\tilde{\mathbf{a}} = \rho \mathbf{a},$$
$$\rho = e^{-2\sigma}, \qquad \sigma \in C^\infty(M), \tag{25}$$

then

$$\delta(\rho a_{\alpha\beta}) = e^{-2\sigma} a_{\alpha\beta} \tilde{\mathrm{w}}, \qquad \tilde{\mathrm{w}} = \mathrm{w} - 2\mathrm{d}\sigma . \tag{26}$$

Hence, if we take $\tilde{\mathbf{a}}$ as the metric tensor instead of **a**, and if at the same time **w** is transformed into $\tilde{\mathbf{w}}$ we get the same covariant derivative. Note that the antisymmetric tensor field

$$F_{\alpha\beta} = \partial_\alpha \varepsilon_\beta - \partial_\beta \varepsilon_\alpha \tag{27}$$

is invariant under this gauge transformation. This tensor field can be interpreted as the one corresponding to the *effective electromagnetic field*. However, this field can be also endowed with the physical meaning of a measure of the influence of the graphene *effective temperature* $\theta$ on the length measurement. Namely, if

$$\varepsilon = \varepsilon(\mathrm{P}(u), \theta, \mathrm{d}\theta) = \varepsilon_\alpha(u, \theta, \mathrm{d}\theta) \mathrm{d}u^\alpha \geq 0 , \tag{28}$$

where $\theta \in C^\infty(M)$, $\theta \geq 0$, is a field of effective absolute temperatures, then since in graphene is observed the *negative thermal expansion*, that is, this material contracts upon heating rather than expands as most materials do (Section 1), this relation should fulfil the following additional condition:

$$\mathrm{d}\theta(u) \geq 0 \Rightarrow 0 \leq \varepsilon_\alpha(u, \theta(u), \mathrm{d}\theta(u)) \leq 1, \qquad \alpha = 1, 2 . \tag{29}$$

It follows from eqs. (17), (22) and (28) that:

**Statement 1** *The Weyl material space admits to define the following relationship between its geometry and an isothermal distribution of its effective absolute temperature $\theta$*:

$$\mathrm{div}_a \mathbf{w} + 2K = r,$$
$$\mathbf{w} = w^\alpha(u, \theta, \mathrm{d}\theta) \partial_\alpha, \tag{30}$$

*where $r \in \mathrm{R}$ is a constant, the condition (29) is assumed,* $\mathbf{w} \in \mathrm{W}(M)$, *and*

$$\mathrm{div}_a \mathbf{w} = \nabla^a_\alpha w^\alpha, \qquad w^\alpha = 2a^{\alpha\beta} \varepsilon_\alpha . \tag{31}$$

Thus, the equation (30) can be considered an *isothermal thermal state equation* describing a relationship between the Gauss curvature of a graphene sheet and its *effective absolute temperature* $\theta$ treated as a *material state parameter*.

If $\varepsilon = \varepsilon(\theta, \mathrm{d}\theta)$ and

$$\varepsilon_\alpha(\theta, \mathrm{d}\theta) = \beta(\theta)\partial_\alpha\theta, \tag{32}$$

then

$$\varepsilon(\theta, \mathrm{d}\theta) = \mathrm{d}\sigma(\theta), \qquad \sigma(\theta) = \int_{\theta_0}^{\theta} \beta(\tau)\mathrm{d}\tau, \tag{33}$$

where, according to eq.(29), the *coefficient of thermal expansion* $\beta(\theta)$ should fulfil the condition

$$\mathrm{d}\theta \geq 0 \Rightarrow 0 \leq \beta(\theta) \leq 1.. \tag{34}$$

In this case

$$\delta(\rho a_{\alpha\beta}) = 0, \tag{35}$$

that is, the *thermal distortion* of the intrinsic length measurement takes the form of a conformal transformation of the intrinsic material metric tensor **a**:

$$\nabla^{a,w} = \nabla^{\tilde{a}}, \tag{36}$$

where eq.(25) was taken into account. Let's notice that because the thermal distortion *of* the internal length measurement is considered and the effective temperature $\theta$ is treated as a material state parameter, so it seems physically reasonable to introduce the *characteristic thermal length parameter l($\theta$)* by the rule

$$\beta(\theta) = l'(\theta)/l(\theta) = \mathrm{d}\ln l(\theta)/l_0, \tag{37}$$

where $l_0 = l(\theta_0)$ is the characteristic length in a reference temperature $\theta_0$. We can say then that the *isothermal material structure* of the graphene sheet is characterized by this characteristic length. In this case eq.(30) reduces to the following relation (see Appendix C):

$$\Delta_a \sigma + K = r/2. \tag{38}$$

Particularly, if

$$K = r = 0, \tag{39}$$

then the material space is *flat* and

$$\Delta_a \sigma = 0. \tag{40}$$

If we assume that $\sigma(\theta)$ depends on the effective temperature $\theta$ linearly, we obtain

$$\Delta_a \theta = 0. \tag{41}$$

This is exactly the same condition encountered in the linear thermoelasticity of isotropic bodies being in isothermal and adiabatic conditions.

It ought to be stressed that the effective temperature $\theta$ is treated here as a material state parameter due to the technology of the production of graphene. Namely, as it was mentioned in Section 1, „the impossibility of growing 2D crystals does not actually mean that they cannot be made artificaly". If so, it ought to be taken into account that, according to the Mermin-Wagner theorem (Section 1; see also e.g., [5]), the existence of corrugated graphene sheets is interconnected with the phenomenon of thermal fluctuations of their lattice atoms. Consequently, we can consider **w** as a *thermal state vector field* or, shortly, as *thermal state vector*.

### 3. Orthogonal configurational spaces

The *ideal graphene sheet* is a subset of a 2-dimensional Euclidean point space (Appendix B) homeomorphic to this space. The *graphene sheet* is a subset of the 3-dimensional point space $A_3$ (treated as a differential manifold - see Appendices A and B) homeomorphic to an ideal graphene sheet. In the paper are considered graphene sheets being *regular surfaces* in the orthogonal point space $A_{3,g}$ (Euclidean or Minkowski type - Appendix B). The regular surfaces in the Euclidean point space will be referred to as *spatial configurations* of graphene. A distinguished spatial configuration is called the *reference configuration* of graphene. Thus, the Euclidean point space $A_{3,g}$ can be named the *configuration space* of graphene sheets (the *mechanical configurational space*). Really observed graphene sheets are their corrugated versions (which physical properties are studied with the help of various 2D geometrical models such as, for example, membranes or plates; Section 1). In the theory of graphene treated as a planar two-dimensional material, the description of its *quantum-mechanical phenomena* needs to consider the graphene sheets embedded (at least locally) in the 3-dimensional Minkowski point space as *space-like regular surfaces* (see Section 1 and Appendix B). In this case we are dealing with the orthogonal geometry describing a *configuration space of internal processes* in a graphene sheet.

Let us consider, as a *global reference configuration* of graphene, a plane $\Sigma_\mathbf{n}$ in the Euclidean point space $A_{3,g}$ normal to a distinguished direction $\mathbf{n} \in W_{3,g}$. We can introduce the *associated Minkowski vector space* $W_{3,h} = (W_3, \mathbf{h})$ defined by the following metric tensor:

$$\mathbf{h} = \mathbf{g} - 2\mathbf{n} \otimes \mathbf{n}, \qquad (\mathbf{n}, \mathbf{n})_g = 1. \tag{42}$$

Since

$$\mathbf{h}|\Sigma_\mathbf{n} = \mathbf{g}|\Sigma_\mathbf{n}, \qquad (\mathbf{n}, \mathbf{n})_h = -1, \tag{43}$$

we conclude that the signature of **h** is (+, +, -). If

$$\mathbf{s} \in \Sigma_\mathbf{n}, \qquad (\mathbf{s}, \mathbf{s})_g = 1, \tag{44}$$

then

$$(\mathbf{s}, \mathbf{n})_h = 0 , \tag{45}$$

and the pair $H_{\mathbf{s},\mathbf{h}} = (H_\mathbf{s}, \mathbf{h}_s)$, where

$$H_\mathbf{s} = \left\{ \mathbf{u} \in W_{3,g} : (\mathbf{u}, \mathbf{s})_g = 0 \right\}, \quad \mathbf{h}_s = \mathbf{h}|H_\mathbf{s}, \tag{46}$$

is called the *hyperbolic plane* and has the signature (+, -).

The vector product $\mathbf{u} \times_h \mathbf{v}$ of vectors belonging to the above defined orthogonal vector space $W_{3,h}$ of Minkowski type can be defined in terms of the vector product $\mathbf{u} \times \mathbf{v}$ in the Euclidean vector space $W_{3,g}$ in the following manner ([29] in the notation of Appendix B):

$$\begin{aligned} \mathbf{u} \times_h \mathbf{v} &= \mathbf{R_n}(\mathbf{u} \times \mathbf{v}), \quad \mathbf{R_n} \in L(W_{3,g}), \\ \mathbf{R_n} &= \mathbf{1} - 2\mathbf{n} \otimes \mathbf{n}, \quad \mathbf{R_n}(\mathbf{u}) = \mathbf{u} - 2(\mathbf{u}, \mathbf{n})_g \mathbf{n}. \end{aligned} \tag{47}$$

Let's notice that the vectors

$$\mathbf{e}_+ = \frac{1}{\sqrt{2}}(\mathbf{s} + \mathbf{n}), \quad \mathbf{e}_- = \frac{1}{\sqrt{2}}(\mathbf{s} - \mathbf{n}), \tag{48}$$

are **h**-isotropic:

$$(\mathbf{e}_+, \mathbf{e}_+)_h = (\mathbf{e}_-, \mathbf{e}_-)_h = 0, \quad (\mathbf{e}_+, \mathbf{e}_-)_h = 1, \tag{49}$$

and

$$\mathbf{s} = \frac{1}{\sqrt{2}}(\mathbf{e}_+ + \mathbf{e}_-), \quad \mathbf{n} = \frac{1}{\sqrt{2}}(\mathbf{e}_+ - \mathbf{e}_-). \tag{50}$$

The question arrives how to define explicitly a coordinate system in the orthogonal configurational space $A_{3,g}$ such that its *coordinate surfaces* are *developable* and normal to a distinguished vector field, i.e., these surfaces are formed by the such *global bending* of a planar coordinate surface that preserves its flatness. Thus, let us consider the orthogonal point space $A_{3,g} = (A_3, W_{3,g}, \varphi)$ (see Appendices A and B). Let $P_0 \in A_3$, $\varphi(P_0) = \mathbf{o} \in W_3$, be the origin of $A_3$, $c \equiv c_\varepsilon = (\mathbf{c}_i)_{1 \leq i \leq 3}$ - the **g**-orthonormal base defined by (173) and (174), and $x = (x^1, x^2, x^3): A_3 \to \mathbb{R}^3$ - the corresponding affine coordinate system defined by eqs.(162) and (163). Let $f: A_3 \to A_3$ be a *global diffeomorphism* and $\mathbf{k}: A_3 \to W_{3,g}$ - a not disappearing nowhere vector field of the class $C^{r-1}, r \geq 2$. The coordinate descriptions $\mathbf{k}_c$ and $f_c$ of mappings $\mathbf{k}$ and $f$ are given by:

$$\begin{aligned} \mathbf{k}_c &= (k_c^1, k_c^2, k_c^3): \mathbb{R}^3 \to \mathbb{R}^3, \\ \mathbf{k} &= k^i \mathbf{c}_i : A_3 \to W_{3,g} \Rightarrow k_c^i = k^i \circ x^{-1} \in C^{r-1}(\mathbb{R}^3), \quad r \geq 2, \\ f_c &= x \circ f \circ x^{-1} = (f_c^1, f_c^2, f_c^3): \mathbb{R}^3 \to \mathbb{R}^3. \end{aligned} \tag{51}$$

If $W_{3,g}$ is the orthogonal Minkowski space (say e.g. endowed with the metric tensor **h** defined by eq.(42)), then we will assume that **k** is a *timelike* vector field (Appendix B). We will denote in this Section by $(\cdot,\cdot)_\varepsilon$, $\varepsilon = \pm 1$, the scalar products in $\mathbb{R}^3$ defined by eq.(174).

**Theorem 1** [34] *Let* $\mathbf{k}_c = (k_c^1, k_c^2, k_c^3): \mathbb{R}^3 \to \mathbb{R}^3$ *be a coordinate description of a vector field* $\mathbf{k}: A_3 \to W_{3,g}$ (*timelike in the case of the orthogonal Minkowski space*), *which satisfies the following conditions:*

(i) $k_c^3(x) \neq 0$ *for each* $x \in \mathbb{R}^3$;

(ii) *there exists a scalar field* $\varphi \in C^r(\mathbb{R}^3)$, $r \geq 2$, *such that*

$$k_c^i(x) = \frac{\partial}{\partial x^i}\varphi(x), \quad i = 1,2,3, \quad \text{for each } x \in \mathbb{R}^3. \tag{52}$$

*Then there exists a coordinate description* $f_c : \mathbb{R}^3 \to \mathbb{R}^3$ *of a diffeomorphism* $f : A_3 \to A_3$ *such that*

$$k_c'^i(x) = \sum_{j=1}^{3} \frac{\partial f_c^i}{\partial x^j} k_c^j(x) = \begin{cases} \sqrt{\left|(\mathbf{k}_c(x), \mathbf{k}_c(x))_\varepsilon\right|}, & \text{if } i = 3, \\ 0, & \text{othervise}, \end{cases} \tag{53}$$

Since **k** is $C^1$, then there exists a *congruence* C[**k**] of curves in $A_3$ generated by **k** and possessing the coordinate representation C[$\mathbf{k}_c$] in $\mathbb{R}^3$ generated by the vector field $\mathbf{k}_c$ of Theorem 1. That is, given any point $x \in \mathbb{R}^3$, there exists a unique integral curve $\sigma_{c,x} \in C[\mathbf{k}_c]$ to which $x$ belongs. For a fixed $x_0 \in \mathbb{R}^3$ let us consider the equation

$$\varphi(x) - \varphi(x_0) = 0 \tag{54}$$

which defines a unique *regular surface* $\Sigma_{c,x_0}$ in $\mathbb{R}^3$. The vector

$$\mathbf{k}_c(x) = \left(\frac{\partial \varphi}{\partial x^i}(x); i \to 1,2,3\right) \in \mathbb{R}^3 \tag{55}$$

is by construction *orthogonal* to the surface in the point $x \in \Sigma_{c,x_0}$. In addition $\sigma_{c,x_0} \in C[\mathbf{k}_c]$ is orthogonal to $\Sigma_{c,x_0}$ in $x_0$. In other words, the congruence C[$\mathbf{k}_c$] is *globally surface orthogonal*.

Let us denote by $x(\mathbf{o}) = o = (0,0,0) \in \mathbb{R}^3$, $\mathbf{o} \in W_{3,g}$, the origin of $x(\mathbf{o}) = o = (0,0,0) \in \mathbb{R}^3$. To construct a new coordinate system in $\mathbb{R}^3$ let us fix a point $o' \in \mathbb{R}^3$ as a new origin and consider $\sigma_{c,o'} \in C[\mathbf{k}_c]$ (with the natural orientation) and $\Sigma_{c,o'}$. Let us consider, in place of the triple of *Carte-* *sian reference* $\left(\mathbb{R}^2, \mathbb{R}, o\right)$ being a coordinate description of the *planar reference configuration* of the graphene considered previously, the triple $\left(\Sigma_{c,o'}, \sigma_{c,o'}, o'\right)$ with $\Sigma_{c,o'}$ being a *regular surface embedded* in $\mathrm{R}^3$ (let's remind that *embedding* is an injective immersion), i.e., $\Sigma_{c,o'}$ is also an 2-dimensional Riemannian manifold. Let us introduce on $\Sigma_{c,o'}$ a coordinate system $u = \left(u^1, u^2\right)$ (notice that this surface has a unique chart in $\mathbb{R}^2$) and let

$$\mathbf{a} = a_{\alpha\beta}\mathrm{d}u^\alpha \otimes \mathrm{d}u^\beta, \quad \alpha, \beta = 1, 2 , \qquad (56)$$

be its metric tensor induced by the Euclidean product in $\mathbb{R}^3$ see Section 4). Then to every point $x \in \mathbb{R}^3$ there corresponds a unique point on $\sigma_{c,o'}$ (intersection between $\sigma_{c,o'}$ and $\Sigma_{c,x}$) with coordinates $u(x) = \left(u^1(x), u^2(x)\right)$ and vice versa. Therefore, the map

$$\begin{aligned} f_c : \mathbb{R}^3 &\to \mathbb{R}^3, \\ x \mapsto \xi = \left(\xi^1, \xi^2, \xi^3\right) &:= f_c(x) = \left(u^1(x), u^2(x), s(x)\right), \end{aligned} \qquad (57)$$

is one to one and possesses a continuous first-order derivative. Notice that it is not necessary to particularize the coordinate *s* to be the length of the arc on $\sigma_{c,o'}$ [34]. In fact we have the following:

**Theorem 2** [34] *Let* $\xi^3 = s$ *be any coordinate on* $\sigma_{c,o'}$.

1) *Introducing on* $\sigma_{c,o'}$ *the positive metric function* $\sigma(s)$, *we have in the case of the Euclidean space* $\mathrm{W}_{3.g}$:

$$k_c'^i(x) = \begin{cases} \sqrt{\dfrac{\left(\mathbf{k}_c(x), \mathbf{k}_c(x)\right)_{+1}}{\sigma(s(x))}}, & \text{if } i = 3, \\ 0, & \text{othervise}, \end{cases} \qquad (58)$$

*and the global coordinate transformation* $f_c : x \mapsto \xi = f_c(x)$, $x \in \mathbb{R}^3$, *transforms the unity tensor* $\delta_{ij}$ (*covering with the standard Euclidean metric tensor in* $\mathbb{R}^3$ - *see eq.(174)) into the metric tensor* $g_{ij}$ *with components*

$$g_{ij}(\xi) = \begin{cases} a_{\alpha\beta}\left(\xi^1, \xi^2\right), & \text{for } i = \alpha, j = \beta;\ \alpha, \beta = 1, 2, \\ 0, & \text{for } i = 3, j = \beta = 1, 2;\ j = 3, i = \alpha = 1, 2, \\ \sigma(\xi^3) & \text{for } i = j = 3. \end{cases} \qquad (59)$$

*The map* $f_c$, *being a global coordinate transformation on the flat space* $\mathbb{R}^3$, *generates the identically vanishing Riemannian curvature tensor associated with* $g_{ij}$.

2) *Introducing on* $\sigma_{c,o'}$ *the negative metric function* $g_{33}(s)$ *, we have in the case of the orthogonal Minkowski space* $W_{3,g}$:

$$k_c'^i(x) = \begin{cases} \sqrt{\dfrac{(\mathbf{k}_c(x), \mathbf{k}_c(x))_{-1}}{g_{33}(s(x))}}, & \text{if } i=3, \\ 0, & \text{othervise}, \end{cases} \quad (60)$$

*and the global coordinate transformation* $f_c : x \mapsto \xi = f_c(x)$ *,* $x \in \mathbb{R}^3$ *, transforms the standard fundamental Minkowskian tensor* $\eta_{ij}$ *(of the signature* (+, +. -) *- see* eq.(174)) *into the metric tensor* $g_{ij}$ *of the same signature and with components*

$$g_{ij}(\xi) = \begin{cases} a_{\alpha\beta}(\xi^1, \xi^2), & \text{for } i=\alpha, j=\beta; \ \alpha, \beta = 1,2, \\ 0, & \text{for } i=3, j=\beta=1,2; \ j=3, i=\alpha=1,2, \\ g_{33}(\xi^3) & \text{for } i=j=3. \end{cases} \quad (61)$$

3) *In both cases, i.e., for the Euclidean as well as for the Minkowski space, the surface* $\Sigma_{c,o'}$ *has the zero Gaussian curvature, i.e., it is a developable surface.*

Finally, if a global coordinate system possesses the developable coordinate surfaces normal to the **k**-direction, then the considered orthogonal metrics can be represented in the following form:

$$\begin{aligned}\mathbf{g}_\varepsilon(\xi) &= g_{ij}(\xi; \varepsilon) d\xi^i \otimes d\xi^j \doteq \mathbf{a}(\xi^1, \xi^2) + g_{33}(\xi^3; \varepsilon) d\xi^3 \otimes d\xi^3, \\ \mathbf{a}(\xi^1, \xi^2) &= a_{\alpha\beta}(\xi^\kappa) d\xi^\alpha \otimes d\xi^\beta, \quad \alpha, \beta, \kappa = 1,2; \ i,j = 1,2,3,\end{aligned} \quad (62)$$

where $[\xi^i] = $ cm and

$$\begin{aligned} g_{33}(\xi^3; +1) &= \sigma(\xi^3) > 0, \\ g_{33}(\xi^3; -1) &= g_{33}(\xi^3) < 0. \end{aligned} \quad (63)$$

In the case of Minkowski space, that is when we are dealing with the configuration space of electronic processes in a graphene sheet, it is considered a *stationary metric* with the *temporal parameter τ* defined by the conditions:

$$\begin{aligned} \xi^3 &= v_F \tau, \quad g_{33}(\xi^3) = -1, \\ [v_F] &= \text{cms}^{-1}, \quad [\tau] = \text{s}, \end{aligned} \quad (64)$$

where $v_F > 0$ is the Fermi velocity (Section 1).

The Minkowski space is an *effective configuration space* of electronic processes in a graphene sheet. If the graphene sheet is corrugated (see Sections 1 and 2), then the non-developable surfaces ought to be taken into account as its configurations (see Section 4) and, correspondingly, a space de-pendent *effective Fermi velocity*, this is let's say $v_F = v_F(\xi^1, \xi^2)$, is then frequently considered (see e.g. [6]). Notice also that the temporal parameter $\tau$ can be treated as an *internal time* not necessary equals to the *dynamical time* appearing when the dynamics (or kinematics) of graphene sheets is considered. The such defined configurational space is not, in general, a flat pseudo-Riemannian space.

A simply connected three-dimensional flat (pseudo-)Riemannian manifold must have a globally defined orthonormal triad of covariant constant vector fields. One can use these vector fields to define globally three coordinate functions which provide a *local isometry* into an orthogonal space (Minkowski or Euclidean space). This is in effect a global version of coordinates that put a flat metric locally in the form of the orthogonal metric. This *local isometry* is called the *developing map*. If the flat manifold is *complete*, then the developing map is a *global isometry* onto the orthogonal space. [9]

Let's consider a smooth moving frame $e = (\mathbf{e}_a)_{1 \leq a \leq 3}$ not generated by a coordinate system and such that (see, for example, Section 5)

$$\mathbf{g} = g_{ab} \mathbf{e}^a \otimes \mathbf{e}^b,$$
$$\operatorname{sgn}(\mathbf{e}_3, \mathbf{e}_3)_g = \varepsilon, \quad \varepsilon = \pm 1, \tag{65}$$

where designations of eqs.(173), (174) and (62) were taken into account, and $e^* = (\mathbf{e}^a)_{1 \leq a \leq 3}$ is the moving coframe dual to $e$. The transformation

$$\mathbf{e}^a \mapsto \mathrm{F}^k = B^k{}_a \mathbf{e}^a,$$
$$\mathbf{B} = \left(B^k{}_a ; \genfrac{}{}{0pt}{}{k \downarrow 1,2,3}{a \rightarrow 1,2,3}\right) : U \subset \mathrm{A}_{3,g} \rightarrow \mathrm{GL}^+(3), \tag{66}$$

defines a moving coframe $F = (\mathrm{F}^k)_{1 \leq k \leq 3}$ generated by a global (local) coordinate system if and only if there exists a global (local) diffeomorphism $\lambda : \mathrm{A}_{3,g} \rightarrow \mathrm{A}_{3,g}$ such that if

$$\mathrm{d}\lambda_\mathrm{P} : \mathrm{T}_\mathrm{P} \mathrm{A}_{3,g} \rightarrow \mathrm{T}_{\lambda(\mathrm{P})} \mathrm{A}_{3,g},$$
$$\mathrm{d}\lambda_\mathrm{P}(\mathbf{u}_\mathrm{P}) = \mathbf{F}(\mathrm{P}) \mathbf{u}_\mathrm{P}, \quad \mathbf{u}_\mathrm{P} \in \mathrm{T}_\mathrm{P} \mathrm{A}_{3,g}, \tag{67}$$

where $\mathrm{T}_\mathrm{P} \mathrm{A}_{3,g} \simeq \mathrm{T}_{\lambda(\mathrm{P})} \mathrm{A}_{3,g} \simeq \mathrm{W}_{3,g}$ (see Appendix B), then

$$\mathbf{F}(\mathrm{P}) = \partial_{k|\lambda(\mathrm{P})} \otimes \mathrm{F}^k(\mathrm{P}). \tag{68}$$

If $(O, x)$ is a coordinate system on $\mathrm{A}_{3,g}$ and

$$\mathrm{F}^k(\mathrm{P}) = F^k{}_m(x(\mathrm{P})) \mathrm{d}x^m_\mathrm{P}, \tag{69}$$

then

$$F^k{}_m = \partial \lambda^k / \partial x^m,$$
$$\lambda^k = x^k \circ \lambda \circ x^{-1} : x(O) \rightarrow \mathbb{R}^3. \tag{70}$$

Particularly, if $\lambda = f$ is the above considered *global bending*, then this diffeomorphism defines the *developing map* that maps the developable coordinate surfaces (space-like if we are dealing with the point space of Minkowski type) into the planar reference configuration $\Sigma_{\mathbf{n}}$.

If $c = (\mathbf{c}_m)_{1 \leq m \leq 3}$ is a global base of the orthogonal vector space $W_{3,g}$ defining a global coordinate system of the orthogonal point space $A_{3,g}$ (Appendix A) such that (cf. eqs.(42) and (43)):

$$\mathbf{c}_1, \mathbf{c}_2 \in \Sigma_{\mathbf{n}}, \quad \mathrm{sgn}(\mathbf{c}_3, \mathbf{c}_3)_g = \varepsilon, \tag{71}$$

$c^* = (\mathrm{c}^m)_{1 \leq m \leq 3}$ is the corresponding dual frame of $W^*_{3,g}$, and

$$\begin{aligned} \mathrm{e}^a &= P^a{}_m \mathrm{c}^m, \\ \mathbf{P} &= \left(P^a{}_m; {}^{a \downarrow 1,2,3}_{m \to 1,2,3}\right): U \subset A_{3,g} \to \mathrm{GL}^+(3), \end{aligned} \tag{72}$$

then, according to eq.(66), we have:

$$F^k{}_m = B^k{}_b P^b{}_m, \tag{73}$$

Let us assume that $S_a = (S, \mathbf{a})$ is a regular surface in the configurational orthogonal point space $A_{3,g}$, e.g. defined by a *local isometric embedding* of a 2-dimensional material space of a corrugated graphene sheet (see Sections 2 and 4, Appendix B and, e.g., [20, 41]), such that components of the metric tensor $\mathbf{a}$ cover with the two-dimensional unity tensor $\delta_{\alpha\beta}$:

$$\mathbf{a} := \delta_{\alpha\beta} \mathrm{e}^\alpha \otimes \mathrm{e}^\beta, \quad \alpha, \beta = 1, 2, \tag{74}$$

and $\mathbf{e}_3$ is the vector field normal to $S_a$. Next, let us consider the tensor $\left(F^k{}_m\right)$ as a *global deformation tensor* of the *planar reference configuration* $\Sigma_{\mathbf{n}}$ of the graphene sheet. The equation (73) can be interpreted then as a decomposition of this deformation tensor on two *local distortions*: the *plastic distortion* $\left(P^b{}_m\right)$ describing a corrugated state of the planar reference configuration of the graphene sheet (and thus defining a parametrization of the corrugated graphene sheet), and the *elastic distortion* $\left(B^k{}_b\right)$ describing a deformation of the corrugated graphene sheet. In our case the plastic distortion is due to the *internal curvature effect* (Sections 1 and 2). Particularly, if the material space of graphene sheets is flat (see eq.(39)), then the field of plastic distortions defines *locally developable* configurations of the graphene sheet diffeomorphic (at least locally) to a graphene sheet located in the planar reference configuration $\Sigma_{\mathbf{n}} \subset A_{3,g}$ (see Sections 4 and 5). It can be, for example, the case of *amorphous graphene* generated by introducing *Stone-Wales defects* into the perfect graphene lattice (Section 1).

## 4. Geometry of embedded graphene sheets

If we are dealing with a *graphene regular surface* in the affine point space $A_{3,g}$ modelled on the orthogonal vector space $W_{3,g}$ (Appendices A and B, Section 3), then the knowledge of its first and second fundamental tensors (defined in this Section) facilitates the analysis of the influence of *surface shape* on the physical states of graphene. First fundamental tensor is an *intrinsic* object of the surface that represents the states of the graphene material structure in a manner invariant with respect to translations and **g**-orthogonal motions of the surface in the ambient 3-space (preserving additionally the space-like type of geometry of the graphene surface if we are dealing with the Minkowski-type point space). Concerning the second fundamental tensor, it is an *extrinsic* tool to characterize the shape of graphene configurations in terms of the orthogonal geometry of the ambient 3-space. It is a basic fact that we can talk about characteristics of the embedded graphene surface measured by means of its second fundamental tensor only when it can be considered as a metric tensor on the surface, that is, only if it is a non-degenerate symmetric 2-covariant tensor field (Appendix B).

So, given a smooth 2-dimensional *Riemannian manifold* $M_a = (M, \mathbf{a})$, say e.g. appearing in the description of the isothermal stability of the graphene material structure due to the *curvature effect* (Section 2). This internal geometry is observed in the Euclidean physical configurational space $A_{3,g}$ as wrinkling the graphene sheet. It is because 2-dimensional Riemannian manifolds are *locally* embeddable in the Euclidean point space $A_{3,g}$ (see remarks below). So, let as assume that there exists a mapping $\kappa: M_a \to A_{3,g}$, $S \equiv M_\kappa := \kappa(M) \subset A_3$, such that $S$ is a *regular surface* (Appendix B) and

$$\kappa^* \mathbf{g} = \mathbf{a}, \tag{75}$$

where **g** is the metric tensor of the *Euclidean point space* $A_{3,g}$, and $\kappa^*$ is the pull back mapping (Appendix C). Let's notice that smoothly and *isometrically* local immersing a 2-dimensional Riemannian manifold $M_a$ into $A_3$ is equivalent to finding *locally* three smooth functions $x^i : M \to \mathbb{R}$, $[x^i]$ = cm (see Appendix C), $i = 1,2,3$, such that

$$\mathbf{a} \doteq \delta_{ij} \mathrm{d}x^i \otimes \mathrm{d}x^j,$$
$$[\mathbf{a}] = \mathrm{cm}^2, \quad [\mathrm{d}x^i] = \mathrm{cm}. \tag{76}$$

In local coordinates $u = (u^1, u^2): O \to \mathbb{R}^2$, $[u^i]$ = cm, $O \subset M$, the metric **a** is of the form

$$\mathbf{a} \doteq a_{\alpha\beta} \mathrm{d}u^\alpha \otimes \mathrm{d}u^\beta, \tag{77}$$

where

$$a_{\alpha\beta} = \delta_{ij} \partial_\alpha x^i \partial_\beta x^j, \quad \partial_\alpha = \partial/\partial u^\alpha, \quad [\partial_\alpha] = \mathrm{cm}^{-1}, \tag{78}$$

$x^i = x^i(u^1, u^2)$, $i = 1,2,3$, are embedding functions and, for the simplicity of the notation, the images

$u(Q) = (u^1(Q), u^2(Q)) \in U \subset \mathbb{R}^2$ of points $Q \in O$ under the mapping $u$, are designated also by $u = (u^1, u^2)$. The mapping $\kappa$ is called *isometric embedding* or *isometric immersion* if $\kappa$ is embedding or immersion, respectively. If $A_{3,g}$ is the Minkowski point space, then above statements are preserved under the condition that the surface $S = M_\kappa$ is a space-like surface (i.e., $S$ is endowed with the positive definite metric tensor [29]). In this case we are saying that $\kappa$ is a *space-like isometric immersion* (or *embedding*) [29]. There are several surveys in this topic (e.g., [12, 20, 29]; Appendix B).

Let $c = (\mathbf{c}_i)_{1 \leq i \leq 3}$, be a **g**-orthonormal base of the orthogonal vector space $W_{3,g}$ fulfilling the condition (71) (i.e., the space can be Euclidean or Minkowski type) with the versor $\mathbf{c}_3$ parallel to the versor **n** appearing in eq.(42). We will denote

$$\mathbf{x} = x^i \mathbf{c}_i : O \subset M \to W_{3,g},$$
$$\mathbf{r} = \mathbf{x} \circ u^{-1} : U \subset \mathbb{R}^2 \to W_{3,g}, \quad u = (u^1, u^2), \tag{79}$$
$$[\mathbf{x}] = [\mathbf{r}] = [1], \quad [\mathbf{c}_i] = \text{cm}^{-1}, \quad [x^i] = [u^\alpha] = \text{cm}.$$

In these designations:

$$a_{\alpha\beta} = (\mathbf{r}_\alpha, \mathbf{r}_\alpha)_g, \quad \mathbf{r}_\alpha = \partial_\alpha \mathbf{r},$$
$$[\mathbf{r}_\alpha] = \text{cm}^{-1}, \quad [a_{\alpha\beta}] = [1]. \tag{80}$$

where $\partial_\alpha = \partial/\partial u^\alpha$, $\alpha = 1, 2$. Let us denote by $S_a = (S, \mathbf{a})$ the range $S = M_\kappa \subset A_{3,g}$ of $M$ under an *isometric regular embedding* $\kappa$ (see Appendix B) endowed with the induced metric **a** defined locally on the surface $S$ by eqs.(77) - (80). The Riemannian 2-dimensional manifold $M_a$ can be *locally identified*, without losing of the generality of description of the internal geometry, with the *regular surface* $S_a$ endowed with the metric **a**. This metric tensor is called then the *first fundamental tensor* of the (regular) surface $S$.

Consider a curve $\gamma : T \to S$ on the surface $S_a$ defined by (see Appendix A)

$$t \in T \subset \mathbb{R} \mapsto \gamma(t) = \overrightarrow{O\gamma(t)} := \mathbf{r}(u(t)), \tag{81}$$

where $t$, $[t] = [1]$, is an affine parameter. The *arc length element* $ds_a$ of this curve is given by

$$ds_a^2(t) = (\dot{\gamma}(t), \dot{\gamma}(t))_a = I(u)(t), \quad [s_a] = \text{cm},$$
$$I(u) = E(u)(du^1)^2 + 2F(u)du^1 du^2 + G(u)(du^2)^2, \tag{82}$$

where

$$E = (\mathbf{r}_1, \mathbf{r}_1)_a, \quad F = (\mathbf{r}_1, \mathbf{r}_2)_a, \quad G = (\mathbf{r}_2, \mathbf{r}_2)_a,$$
$$[\dot{\gamma}] = [E] = [F] = [G] = [1], \quad \dot{\gamma}(t) = d\gamma(t)/dt, \tag{83}$$

and subscripts 1, 2 denote partial derivatives. $I$, $[I] = cm^2$, is called the *first fundamental form*. One then introduces embedding functions $x^i = x^i(u)$, $[x^i] = cm$, $i = 1,2,3$, such that

$$\delta_{ij} dx^i(u) dx^j(u) = I(u), \tag{84}$$

which implies

$$E = \delta_{ij} x_1^i x_1^j, \qquad F = \delta_{ij} x_1^i x_2^j, \qquad G = \delta_{ij} x_2^i x_2^j, \tag{85}$$

where $x^i_\alpha = \partial x^i / \partial u^\alpha$. Notice that if $x' = (x'^i; i \to 1, 2, 3): U \subset \mathbb{R}^2 \to \mathbb{R}^3$ is the different embedding mapping that defines the same first fundamental form $I(u)$, then the so-called *Rigidity Theorem* [22] states that the mapping $x' \circ x^{-1}: x(U) \to x'(U)$ is an isometry in $\mathbb{R}^3$.

The system (85) of non-linear first order partial differential equations is not any standard type. For example, it is known, that any analytic 2-dimensional Riemannian manifold admits a *local analytic* isometric embedding in $\mathbb{R}^3$ (and thus - in $A_{3,g}$) [20, 22], while any smooth 2-dimensional Riemannian manifold admits a local smooth isometric embedding in $\mathbb{R}^4$ [22]. However, for any point of a $C^1$ 2-dimensional Riemannian manifold there is a neighbourhood which has a $C^1$ isometric embedding in $\mathbb{R}^3$ [12] and any smooth *nonnegatively curved* metric always admits a local smooth embedding in $\mathbb{R}^3$ [22]. Moreover, if the Gaussian curvature $K$ of a smooth surface $S_a$ satisfies the following condition at the point $\mathrm{P} \in S_a$

$$K(\mathrm{P}) = 0 \text{ and } dK(\mathrm{P}) \neq 0, \tag{86}$$

or it satisfies the condition

$$dK(\mathrm{P}) \leq 0 \text{ and } d^2 K(\mathrm{P}) \neq 0, \tag{87}$$

then it admits a smooth local isometric embedding in $\mathbb{R}^3$ near P [22].

Let's notice that, according to Theorem 3, the existence of a *global isometric embedding* needs the existence of a *global planar reference configuration* (see Section 3). Moreover, the following theorems hold:

**Theorem 3** (Embeddings) [22]
**3.1** (Olovjasnikov-Pogorelov) *Any smooth complete positive curvature metric defined on* $\mathrm{R}^2$ *admits a smooth global isometric embedding in* $\mathbb{R}^3$.
**3.2** (Hilbert-Efimov) *Any complete surface with negative constant curvature* (*with curvature bounded above by a negative constant*) *has no smooth global isometric immersion in* $\mathbb{R}^3$.

The Gaussian curvature $K$ characterizes the intrinsic geometry of the surface. But for a complete description of its embedding into the Euclidean point space $A_{3,g}$ we need additionally to introduce the so-called *second fundamental tensor* of $S_a = M_{\kappa,a}$:

$$\mathbf{b} = b_{\alpha\beta} du^\alpha \otimes du^\beta, \quad [\mathbf{b}] = \text{cm},$$
$$b = \det(b_{\alpha\beta}), \quad [b] = \text{cm}^{-2}, \tag{88}$$

where

$$b_{\alpha\beta}(u) = (\partial_\alpha \mathbf{r}_\beta, \mathbf{n})_g, \quad [b_{\alpha\beta}] = \text{cm}^{-1}, \tag{89}$$

and further on the vector field

$$\mathbf{n} = \frac{\mathbf{r}_1 \times \mathbf{r}_2}{\|\mathbf{r}_1 \times \mathbf{r}_2\|_g}, \quad [\mathbf{n}] = \text{cm}^{-1}, \tag{90}$$

denotes the unit vector field normal to the surface. It is an *external tensorial measure* of the surface geometry. For example, it is known that the second fundamental tensor of a surface in the Euclidean point space $A_{3,g}$ is non-degenerate (i.e., $b \neq 0$) if and only if this surface is *non-developable* [35]. The same statement is valid in the case of a space-like surface in the affine point space modeled on the orthogonal Minkowski space. Embedding of the surface into $A_{3,g}$ is dictated now by the *Gauss-Weingarten equations* [8]:

$$b_{\alpha\beta}\mathbf{n} = \partial_\alpha \mathbf{r}_\beta - \Gamma^\kappa_{\alpha\beta}[\mathbf{a}]\mathbf{r}_\kappa,$$
$$\partial_\alpha \mathbf{n} = -a^{\beta\kappa} b_{\alpha\kappa} \mathbf{r}_\beta, \quad \alpha, \beta, \kappa = 1,2, \tag{91}$$

where $\Gamma^\kappa_{\alpha\beta}[\mathbf{a}]$ are the Christoffel symbols corresponding to the first fundamental tensor $\mathbf{a}$.

In order to quantify the curvatures of the surface $S_a \subset A_{3,g}$, we consider the so called *natural curve* $\gamma$ on $S_a$ defined by the condition that $\mathbf{l} = \dot{\boldsymbol{\gamma}}$, $[\mathbf{l}] = \text{cm}^{-1}$, is its unit tangent [41]. Let $\boldsymbol{\kappa}$ denotes the *curvature vector* of this curve [17, 36]:

$$\boldsymbol{\kappa} = \frac{d\mathbf{l}}{ds} = \boldsymbol{\kappa}_a + \boldsymbol{\kappa}_n, \quad (\boldsymbol{\kappa}_a, \boldsymbol{\kappa}_n)_g = 0,$$
$$\boldsymbol{\kappa}_n = \kappa_n \mathbf{n}, \quad \kappa_n = (\mathbf{l}, \mathbf{l})_b = b_{\alpha\beta} l^\alpha l^\beta, \quad [\kappa_n] = \text{cm}^{-1}, \tag{92}$$

where $s$, $[s] = \text{cm}$, is the length of the arc on $\gamma$ and $(\mathbf{u}, \mathbf{v})_b$ denotes the bilinear symmetric form defined by the second fundamental tensor $\mathbf{b}$. The vector $\boldsymbol{\kappa}_n$ is called the *normal curvature vector* of the surface $S_a$ in the direction $\mathbf{l}$ and the scalar $\kappa_n$ is called the *normal curvature* (of the surface $S_a$ in the direction $\mathbf{l}$). The *second fundamental form* $II = II(u)$ is defined by

$$II := b_{\alpha\beta} du^\alpha du^\beta = L(du^1)^2 + 2M du^1 du^2 + N(du^2)^2, \tag{93}$$

where $[II] = \text{cm}$, and [56]

$$L = (\mathbf{n}, \partial_1 \mathbf{r}_1)_g = -(\mathbf{r}_1, \partial_1 \mathbf{n})_g, \quad N = (\mathbf{n}, \partial_2 \mathbf{r}_2)_g = -(\mathbf{r}_2, \partial_2 \mathbf{n})_g,$$
$$M = (\mathbf{n}, \partial_2 \mathbf{r}_1)_g = -(\mathbf{r}_1, \partial_2 \mathbf{n})_g = -(\mathbf{r}_2, \partial_1 \mathbf{n})_g, \qquad (94)$$

where $[M] = [N] = [L] = \text{cm}^{-1}$. Let us denote

$$\varphi = u^2 \circ (u^1)^{-1} : u^1(T) \to \mathbb{R},$$
$$T = \langle 0, \ l \rangle, \qquad \lambda = d\varphi/du^1, \qquad (95)$$

where $u^1 : s \in T \mapsto u^1 \equiv u^1(s) \in \mathbb{R}$. The normal curvature can be expressed by [56]:

$$\kappa_n = \frac{II}{I} = \frac{N\lambda^2 + M\lambda + L}{G\lambda^2 + F\lambda + E}. \qquad (96)$$

The extreme value of $\kappa_n$ can be obtained by evaluating the condition $d\kappa_n/d\lambda = 0$ of eq.(96), which gives the condition [56]:

$$\kappa_n^2 - 2H\kappa_n + K = 0, \qquad (97)$$

where

$$K = \frac{LN - M^2}{EG - F^2}, \quad H = \frac{EN + GL - 2FM}{2(EG - F^2)}. \qquad (98)$$

If

$$H^2 \geq K, \qquad (99)$$

then the above equation defines the so-called *principal curvatures* $\kappa_1$ and $\kappa_2$, $\kappa_1 \geq \kappa_2$, such that

$$K = \kappa_1 \kappa_2, \qquad H = \frac{1}{2}(\kappa_1 + \kappa_2), \qquad (100)$$

or equivalently [8, 43]:

$$K = \det(\mathbf{a}^{-1}\mathbf{b}) = b/a \doteq \det(b^\alpha{}_\beta),$$
$$2H = \text{tr}(\mathbf{a}^{-1}\mathbf{b}) \doteq \text{tr}(b^\alpha{}_\beta), \qquad b^\alpha{}_\beta = a^{\alpha\kappa} b_{\kappa\beta}. \qquad (101)$$

From eqs.(89) and (101) follows

**Statement 2** *A non-developable surface $S_a \subset A_{3,g}$ with the positive Gaussian curvature K posseses a positive definite second fundamental tensor $\mathbf{b}$ if $S_a$ is appropriately oriented.*

Let's notice that, according to our convention concerning the dimensions of geometric objects (see Appendix C, eqs.(76) - (80), and eqs.(88) - (90)), the second fundamental tensor can be treated as a measure of the length only after its *rescaling*, say, for example, in this manner:

$$\mathbf{b} \to \mathbf{b}_\theta = l(\theta) \mathbf{b}, \quad [l(\theta)] = \text{cm}, \qquad (102)$$

where $l(\theta)$ is the *characteristic thermal length* at the temperature $\theta$ (Section 2).

So, if a graphene sheet is embedded in a configurational space (its space of configurations or its space of internal processes - see definitions on the beginning of Section 3), then its physical properties can be represented not only by its *topological shape effects* (say e.g. the difference of physical properties of fullerenes and nanotubes) but also by its *non-topological shape effects* depended on both its curvatures (Gaussian and mean). The case of *developable* surfaces was discussed in Section 3. If the surfaces are *non-developable*, then the problem of the formulation of physically sensible *constitutive relation* between the energy and curvatures leads us to the theory of *regular surfaces* with the geometry obtained from a *variational principle*. For this purpose, let us consider a variational principle based on the *energy functional* $\mathbb{E}$ which is defined for open regular domains $O \subset S_a$ of the regular surface $S_a$ in the Euclidean point space $A_{3,g}$ (Appendix B) by:

$$\mathbb{E}(O; H, K) = \int_O e(H, K) \, dS,$$
$$\det \mathbf{b} \doteq \det(b_{\alpha\beta}) \neq 0, \tag{103}$$

where the *surface energy density* $e$ is some function of the curvatures $H$ (mean) and $K$ (Gauss) of the (regular) surface $S_a$. The first variation of the functional $\mathbb{E}$ gives the following *Euler-Lagrangian equation* for the energy density function $e$ ([43]; see also references therein):

$$\frac{1}{2}\Delta_a(\partial e/\partial H) + \Lambda_{a,b}(\partial e/\partial K) + (2H^2 - K)(\partial e/\partial H) + 2KH(\partial e/\partial K) - 2He = 0, \tag{104}$$

where the differential operators $\Delta_a$ and $\Lambda_{a,b}$ are defined as (see Appendix C)

$$\Delta_a = \frac{1}{\sqrt{a}}\frac{\partial}{\partial u^\alpha}\left(\sqrt{a}\, a^{\alpha\beta}\frac{\partial}{\partial u^\beta}\right), \qquad \Lambda_{a,b} = \frac{1}{\sqrt{a}}\frac{\partial}{\partial u^\alpha}\left(\sqrt{a}\, K b^{\alpha\beta}\frac{\partial}{\partial u^\beta}\right), \tag{105}$$

$a^{\alpha\beta}$ and $b^{\alpha\beta}$ are the inverse components of the first and second fundamental tensors, respectively, and $a = \det \mathbf{a} \doteq \det(a_{\alpha\beta})$.

If the Riemannian geometry of the graphene sheet is consistent with the Weyl material space defined in Section 2, that is the conditions (75) - (80) are fulfilled and thus the Riemannian material space $M_a$ can be locally identified with the regular surface $S_a$, then we can introduce the induced „*thermal state vector*" $\mathbf{v}$ (see the commentary at the end of Section 2) as

$$\mathbf{v} = \kappa_* \mathbf{w} \in W(S_a), \tag{106}$$

and eq.(30) transforms oneself onto the following isothermal *thermal state equation*:

$$\text{div}_a \mathbf{v} + 2K = r,$$
$$\mathbf{v} = v^\alpha(u, \theta, d\theta)\partial_\alpha, \qquad v^\alpha = 2a^{\alpha\beta}\varepsilon_\alpha, \tag{107}$$

where the differential operator $\text{div}_a$ is defined by eq.(31), the covector field $\varepsilon$ is defined by eq.(28) for a temperature field $\theta$ defined on the surface $S_a$, and the condition (29) should be taken into account. If the *material Riemannian manifold* $M_a = (M, \mathbf{a})$ is flat, then the observed graphene sheet is *locally developable* (see Section 3) and the condition (107) with $K = 0$ ought to be taken into account. If the graphene sheet is a *non-developable* surface, then the condition (104) ought to be additionally fulfilled. The system of equations (104) - (107) defines then the geometry of *isothermal configurations* of the nonplanar corrugated graphene sheet represented by the pair $(\mathbf{a}, \mathbf{v})$. This pair defines also the local description of *isothermal material Weyl geometry* discussed in Section 2.

Particularly, if the case defined by eqs.(32) - (34) and (37) is considered, then the material Weyl geometry of the embedded corrugated graphene sheet reduces to the conformal rescaling $\mathbf{a} \to \tilde{\mathbf{a}} = \mathbf{a}_\theta$, where

$$\mathbf{a}_\theta = e^{-2\sigma(\theta)}\mathbf{a} = \eta(\theta)^2 \mathbf{a}, \qquad \eta(\theta) = l_o / l(\theta), \tag{108}$$

and $\theta: S \to \mathrm{R}_+$ is a distribution of absolute temperature. The Gaussian curvature transforms according to the rule $K \to \tilde{K} = K_\theta$, where

$$K_\theta = e^{2\sigma(\theta)}\left(\Delta_a \sigma_\theta + K\right), \tag{109}$$

and it was denoted $\sigma_\theta = \sigma(\theta) \equiv \sigma \circ \theta$. Thus, according to eq.(38), we obtain that

$$K_\theta = \frac{r}{2}e^{2\sigma(\theta)} = \frac{r}{2}\eta(\theta)^{-2}, \qquad [r] = \mathrm{cm}^{-2}. \tag{110}$$

It means that, in this particular case (see Theorem 3), the Gaussian curvature can not change the sign. If $r \neq 0$, then a *spherical* or *saddle-shaped* surface occurs alone what corresponds to the case when the pentagon or heptagon defect occurs alone, respectively; if $r = 0$, then, for example, the case of a developable *amorphous graphene* sheet can takes place (Section 1).

Let's notice that the existence of the characteristic thermal length parameter $l(\theta)$ (see eqs.(37) and (108)) enables to introduce the following non-topological *dimensionless thermal shape parameter* of non-developable graphene sheets:

$$\upsilon(\theta) := \frac{H}{l(\theta)K} \tag{111}$$

representing the observed correlation of the curvature of surface with the thermal state of this surface (Section 1).

## 5. Congruence of lines generated by the thermal state vector field

The proposed model of isothermal geometry of corrugated graphene sheets is associated with two distinguished vector fields. Firstly, the coupling of the curvature and thermal effects is represented by the thermal state vector $\mathbf{v}$ tangent to the sheet (eqs.(106), (107) and Section 2). Secondly, a vector field normal to the sheet appears (Section 3 - the vector field **k** and Section 4 - the unit vector field **n** normal to the regular surface $S_a$). Suppose that the vector field **v** disappears nowhere and denote

$$\mathbf{l} = \frac{\mathbf{v}}{\|\mathbf{v}\|_g} . \tag{112}$$

Now, we can consider the ordered triple (**l**, **m**, **n**) of **g** - orthonormal smooth vector fields tangent to the configurational orthogonal point space $A_{3,g}$ and such that the pair (**l**, **m**) is a base of the module $W_\mathbf{n}(A_{3,g})$ of all smooth vector fields normal to the **n** - direction (that is, tangent to the surface $S_a$). Let us remind that if the space $A_{3,g}$ is Minkowski type, then $S_a$ is a space-like surface and **n** - a timelike vector field. Next, let us assume that this triple covers with the *Frenet moving frame* of the congruence C[**l**] of lines generated by the direction **l**. The Frenet moving frame is defined by the *generalized formulas of Frenet* (called also the *Serret-Frenet formulas*), which may serve as a definition of the curvature κ and torsion τ of lines of the congruence C[**l**] [29, 32, 46]:

$$\begin{aligned}
\boldsymbol{\kappa} &= \nabla^g_\mathbf{l} \mathbf{l} = \kappa \mathbf{n}, & \kappa &> 0, \\
\nabla^g_\mathbf{l} \mathbf{n} &= -\varepsilon \kappa \mathbf{l} + \tau \mathbf{m}, & \tau &\geq 0, \\
\nabla^g_\mathbf{l} \mathbf{m} &= -\varepsilon \tau \mathbf{n}, & \kappa, \tau &\in C^\infty(A_{3,g}),
\end{aligned} \tag{113}$$

where

$$\varepsilon = \begin{cases} +1 & \text{Euclidean space,} \\ -1 & \text{Minkowski space,} \end{cases} \tag{114}$$

and in the case of Minkowski space (of the signature (+,+,-)) are considered space-like curves (i.e., the *tangent* **l** and the *binormal* **m** are space-like vector fields); the *principal normal* **n** is then a timelike vector field. The principal normal is indeterminate when the curvature κ vanishes. The Frenet moving frame defines three distributions of planes: $\pi_\mathbf{m} = \pi(\mathbf{l}, \mathbf{n})$ - *osculating planes*, $\pi_\mathbf{l} = \pi(\mathbf{n}, \mathbf{m})$ - *normal planes*, $\pi_\mathbf{n} = \pi(\mathbf{m}, \mathbf{l})$ - *rectifying planes*. The vector field **κ** is the *curvature vector* of the congruence C[**l**].

Let us restrict ourselves to the case of the Euclidean configurational space $A_{3,g}$. Then, in the case of *nonplanar* graphene sheets, the rectifying planes are tangent to the non-developable regular surfaces (Section 4). In the case of *flat* graphene sheets the rectifying planes are tangent to developable surfaces (Section 3).

We can define, for a curvilinear coordinate system on $A_{3,g}$ (or, more generally, in the case of a general three-dimensional Riemannian space endowed with the metric tensor **g**), the so called *curl operator* of vector fields $\mathbf{v} = v^k \partial_k$ tangent to this manifold, by

$$\begin{aligned}
\mathbf{u} &= \text{curl}\,\mathbf{v} = u^k \partial_k, \\
u^k &= e^{klm} \nabla^g_l v_m = e^{klm} \nabla^g_{[l} v_{m]},
\end{aligned} \tag{115}$$

where $\nabla^g$ is the Levi-Civita covariant derivative corresponding to the metric tensor **g**, $e^{klm}$ is the so called *Ricci vector* defined as

$$e^{klm} = \sqrt{g}\varepsilon^{klm}, \qquad g = \det(g_{kl}), \tag{116}$$

and $\varepsilon^{klm}$ is the permutation symbol. This operator is frequently written in the form

$$\mathrm{curl}\,\mathbf{v} = \nabla \times \mathbf{v} \tag{117}$$

which has its origin in the Euclidean cross product:

$$\mathbf{c} = \mathbf{a} \times \mathbf{b} \Leftrightarrow c^k = g^{kl}\varepsilon_{lmn}a^m b^n, \tag{118}$$

where $g_{kl} = (\mathbf{e}_k, \mathbf{e}_l)_g$, and $(\mathbf{e}_k)_{1 \le k \le 3}$ is a vector base of the Euclidean vector space $W_{3,g}$.

Let us quote some formulas and statements taken from the paper [32]. It follows from eqs.(113) - (115) that curl**l** has no components in the principal normal:

$$\mathrm{curl}\,\mathbf{l} = \omega_l \mathbf{l} + \kappa \mathbf{m}. \tag{119}$$

It tell us that if a surface were to exist which contained both **l**-lines and **m**-lines, then **l**-lines would have to be geodesics on this surface, and **m**-lines would have to be parallel on the surface. Let us now suppose that at each point P of $A_{3,g}$ the triad (**l**, **m**, **n**) be such that

$$(\mathbf{n}, \mathrm{curl}\,\mathbf{n})_g = 0. \tag{120}$$

Since the condition (120) is necessary and sufficient for the existence of a family of surfaces whose normal coincide with the principal normal of **l**-lines, we have

**Theorem 4** [32] *The lines of the congruence* C[**l**] *will be geodesics on a family of surfaces, if and only if the condition* (120) *holds. Moreover, the condition* (120) *is equivalent to the condition*

$$\mathbf{n} = \psi\,\mathrm{grad}\,\varphi, \tag{121}$$

*and the condition* (121) *is necessary and sufficient for the existence of a family surfaces*

$$\Sigma_c = \varphi^{-1}(c) \subset A_{3,g}, \qquad c = \mathrm{const.} \tag{122}$$

*whose normal is* **n**.

Let us consider the representation of the Weyl material space by the pair $(S_a, \mathbf{v})$ located in the Euclidean point space $A_{3,g}$ (Sections 2 and 4) and such that the surface $\Sigma_c$ is a slice of the regular surface $S_a$. Since the tangent planes to the surface $\Sigma_c$ are spanned by **l** and **m**, their integral curves are contained in this surface. Thus, although in general the arc distances along an **n**-line can not be considered as coordinates, we may consider coordinates $u = (u^1, u^2)$ on $\Sigma_c$ such that $u^1$-parametric curves ($u^2 = m$ = constant) cover with the **l**-lines and such that $u^2$-parametric curves ($u^1 = l =$ constant) cover with the **m**-lines). Moreover, according to the above theorem, we can take for the coordinate $u^1$ the length $l$ of arc of the geodesics of $\Sigma_c$ tangent to the **l**-direction, that is the metric tensor of $\Sigma_c$ is [32]

$$\mathbf{a} = du^1 \otimes du^1 + a(u^1, u^2)\,du^2 \otimes du^2. \tag{123}$$

The *mean curvature H* and the *Gaussian curvature K* of the such defined surface $S_a$ are (in designations of eq.(100); in [32] it is assumed that $H = \kappa_1 + \kappa_2$ )

$$2H = -\text{div}\mathbf{n},\qquad(124)$$

and

$$K = -\kappa(\kappa + \text{div}\mathbf{n}) - \tau^2 .\qquad(125)$$

It follows from eqs.(124) and (125) that the curvature $\kappa$ and the torsion $\tau$ appearing in the generalized formulas of Frenet and the curvatures of the considered regular surface are coupled by the relation

$$2\frac{H}{\kappa} - \frac{K}{\kappa^2} = 1 + \left(\frac{\tau}{\kappa}\right)^2 .\qquad(126)$$

Thus, if the condition (120) is fulfilled, then apart of the condition (107) which limits the thermal state versor **l** of eq.(112), the relation (126) ought to be fulfilled.

Let's notice that the *Darboux vector*

$$\mathbf{d} = \tau\mathbf{l} + \kappa\mathbf{m}\qquad(127)$$

is inclined to direction **l** at the angle $\vartheta$, where [32]

$$\text{ctg}\,\vartheta = \frac{\tau}{\kappa} .\qquad(128)$$

Thus, according to eqs.(111), (126) and (128), the thermal shape parameter $\upsilon(\theta)$ of the non-developable regular surface $S_a$ such that $\Sigma_c \subset S_a$ has the following representation:

$$\upsilon(\theta) = \frac{1}{2l(\theta)}\left[\frac{1}{\kappa} + \frac{\kappa}{K}\left(1 + \text{ctg}^2\vartheta\right)\right] .\qquad(129)$$

The **m**-lines can be geodesics only if

$$K = 0,\qquad(130)$$

that is only if the surface $\Sigma_c$ of eq.(122) is *developable*. It can be shown that the only type of surface that can accommodate orthogonal families of geodesics is a developable [32]. For example, if , in addition to the condition (120), the vector field **l** satisfies the following condition:

$$\begin{aligned}
&\left(\text{grad}\mathbf{l}(\mathbf{m}),\,\mathbf{m}\right)_g = 0,\\
&\text{grad}\mathbf{v}(\mathbf{u}) := \nabla_{\mathbf{u}}^g \mathbf{v},\qquad \mathbf{u} \in W(A_{3,g}),\\
&\text{grad}\mathbf{v} \in W(A_{3,g}) \otimes W(A_{3,g})^*,\qquad \mathbf{v} \in W(A_{3,g}),
\end{aligned}\qquad(131)$$

where the Euclidean point space $A_{3,g}$ is treated as a Riemannian manifold (Appendix B), then the surface $\Sigma_c$ is a *developable*. From the condition (112) it follows that then the thermal state vector **v** has the form

$$\mathbf{v} = v\,\mathbf{l},\qquad(132)$$

where **l** is the versor which appears in the generalized of formulas of Frenet and fulfils the conditions (120) and (131). In this case the vector field **v** can be named, modify the onomastic of [32], a *flat thermal state vector*. The first fundamental form of $\Sigma_c$ associated with the flat thermal state vector is

$$ds_a^2 = dl^2 + dm^2. \tag{133}$$

It means that the **l**-lines and the **m**-lines comprise a two dimensional Euclidean space on this surface (see, for example, the *Stone-Wales* defect; Section 1). The conditions (107) and (126) reduce then to

$$\text{div}_a \mathbf{v} = r, \quad r = constant, \tag{134}$$

and

$$2\frac{H}{\kappa} = 1 + \left(\frac{\tau}{\kappa}\right)^2. \tag{135}$$

respectively.

The *rectifying plane* belonging to the distribution $\pi_\mathbf{n}$ of planes has the equation

$$\left(\overline{P(l)X}, \mathbf{n}(l)\right)_g = 0 \tag{136}$$

where X is a point on the plane and P($l$) is a point on the **l**-line (with its length of the arc parameter $l \geq 0$, [$l$] = cm). As $l$ varies the rectifying planes comprise a one-parameter family of planes which envelope the *rectifying developable* of the **l**-line. We have (cf. Theorem 2):

**Theorem 5** [32] *A normal congruence of developable surfaces defined by the condition* (120) *is possible if and only if a representative surface* $\varphi(P) = constant$ *is the common rectifying developable of all **l**-lines situated upon it.*

Thus, the representative surface $\varphi(P) = constant$ is then a locally developable surface endowed with the flat thermal state vector (see Section 3 - locally developable surfaces defined by local plastic distortions). The thermal state **l**-line is a curve upon this surface. By Theorem 4 it is a geodesic on the surface. Finally, we see that the Frenet moving frame (**l**, **m**, **n**) of eq.(113) with $\varepsilon = 1$ defines an orthogonal family of geodesics on the graphene sheet $S_a$ with its normal versor field **n** if and only if the slice $\Sigma_c = \varphi^{-1}(c)$ of this sheet is a developable surface. It is the case of the *flat Riemannian material space* $M_a = (M, \mathbf{a})$ (Section 2) with the parametrization of its local isometric embedding in $A_{3,g}$ (Section 4) defined by the plastic distortions (Section 3) and endowed with the *flat thermal state vector* defined by eqs.(112) and (132) - (135).

## 6. Conclusions and remarks

The new method of description of isothermal corrugations of graphene sheets proposed in the paper is based on the notion of the *Weyl material space* introduced in Section 2. This theory begins with the observation that the curvature and the lattice thermal fluctuations of graphene sheets are coupled with their thermal stability (see Section 1). It is a counterpart of the *Tolman-Ehrenfest relation* considered in the theory of stationary space-times (e.g. [37]). Namely, this relation states that the temperature is not constant in space at equilibrium but varies with its curvature. However, in contrast to the case of Tolman-Ehrefest relation, the 2D-dimensional Weyl material space can be observed (at least locally) in the (2D+1)-dimensional physical configurational space (Sections 3 - 5). Consequently, a graphene sheet can be treated as a thermodynamic system in diathermal and isothermal conditions (cf. [47] and [48]) which is in contact with a heat bath (e.g. [18, 21]). Next, a graphene sheet is additionally endowed with the distinguished thermal state vector field defined by eqs.(106) and (107). It enables to consider the equation (107) and the condition (126) as relations that can be associated with the influence of the heat bath on this sheet (cf. [37]).

The proposed model of the isothermal geometry of graphene sheets concerns, in general, the case of local continuous thermomechanic description of macroscopic graphene samples (see Section 1 and the estimation for planar samples quoted therein). However, the model of a nonplanar regular surface defined by eq.(104) means that the influence of the mean curvature $H$ on the surface energy of graphene sheets is taken into account. Consequently, effects conditioned not only by the intrinsic geometry of a graphene sheet (represented by the Gauss curvature $K$ - Section 2) but also by his shape observed in the configurational space (Sections 4 and 5), can be taken into account (see remarks in Section 1 concerning the non-local effects and, for example, [2, 47, 48]). For example, the existence of the thermal state vector field and the characteristic thermal length parameter (Section 2) enables to introduce a common representation of these effects in the form of the non-topological dimensionless *thermal shape parameter* (eq.(111)) which is dependent on the geometry of the congruence of lines generated by the thermal state vector field (eq.(129)) and is consistent with the observed correlation of the geometry of graphene sheets with their thermal state (Section 1).

Let us compare the proposed description of corrugated graphene sheets as regular surfaces with the variational geometry based on the curvature dependent surface energy density (Section 4) with the model of the surface tension proposed in papers [1] and [2]. It was observed in [1] that the *membrane model* of the surface tension is not adequate for the satisfactory description of mechanical behaviour of such an interface for which the energy density depends on curvatures only (see remarks in Section 1 relating to this problem). For this reason, a more complex *Cosserat two-dimensional model* (*shell model*) (see e.g. [19]) was proposed in [1]. However, this approach leads to the more general theory of material surfaces than this is necessary to the description of curvature dependent surface tension only [1, 2]. For example, the case of boundary surfaces of incompressible fluids endowed with the surface energy density of form $e(H,K) = aH + bK + c$, where $a$, $b$, $c$ are some constants, was considered in [2]. In this case, the general approach presented in [1] (and reformulated in [2]) leads, when the surface cannot transmit tangential forces, to the following generalization of the well known Laplace formula: $(\mathbf{p}, \mathbf{n})_g = \alpha H + \beta K$, where $\alpha$, $\beta$ are constants, $\mathbf{n}$ is the unit vector normal to the surface and $\mathbf{p}$ is the vector of the surface density of external forces acting on the surface (it can be e.g. the resultant of forces exerted on the surface of separation of two liquid phases). It suggests that the 2D-model of Cosserat continuum can be useful to the description of a broad class of internal forces acting in corrugated graphene sheets treated as solid surfaces endowed with the variational isothermal geometry (see remarks in Section 1 concerning this problem). The differential equation (104) can be treated then as a *constitutive relation* that defines a relationship between the energy density and curvatures of the considered Cosserat surface. The equations (107), (112) and (126) couple then the geometry of this surface with the isothermal distribution of its temperature. Consequently, this approach enables to consider a relationship between the Cosserat-type internal forces of a corrugated graphene sheet and its temperature. We leave the detailed analysis of this topic to a separate study.

**Appendix A - Affine spaces and mappings** [38, 42]

Let $W_n$ be the n-dimensional real vector space, $L(W_n)$ - the real linear space of all R-linear mappings $L: W_n \rightarrow W_n$, and let $T_1^1(W_n)$ denotes the real linear space of all 1-contravariant and 1-covariant tensors over the vector space $W_n$. If $e = (\mathbf{e}_i)_{1 \leq i \leq n}$ is an ordered base in $W_n$ and $e^* = (e^i)_{1 \leq i \leq n}$ is the base of the covector space $W_n^*$ dual to $e$, then the tensor

$$\mathbf{A} = A_i^{\ k} e^i \otimes \mathbf{e}_k \in T_1^1(W_n) = W_n^* \otimes W_n,$$
$$\langle e^i, \mathbf{e}_k \rangle \equiv e^i(\mathbf{e}_k) = \delta_k^i, \tag{137}$$

acts according to the following rules:

$$\mathbf{x} = x^i \mathbf{e}_i \in W_n \Rightarrow \mathbf{y} = \mathbf{A}\mathbf{x} = A_i^{\ k} x^j \langle e^i, \mathbf{e}_j \rangle \mathbf{e}_k,$$
$$\mathbf{y} = y^k \mathbf{e}_k \in W_n, \qquad y^k = A_j^{\ k} x^j, \tag{138}$$

and if additionally

$$\mathbf{x} = \mathbf{B}\mathbf{u}, \qquad \mathbf{B} = B_k^{\ i} e^k \otimes \mathbf{e}_i, \qquad \mathbf{u} \in W_n, \tag{139}$$

then

$$\mathbf{y} = \mathbf{C}\mathbf{u}, \qquad \mathbf{C} = \mathbf{A}\mathbf{B} = C_j^{\ i} e^j \otimes \mathbf{e}_i, \qquad C_j^{\ i} = A_k^{\ i} B_j^{\ k}. \tag{140}$$

The above formula, treated as a multiplication rule in the $\mathbb{R}$ - linear space $T_1^1(W_n)$, defines in this space the structure of a ring with the unit element $\mathbf{1} = \delta_j^i \mathbf{e}^j \otimes \mathbf{e}_i$, where $\delta_j^i \equiv \delta_j^i$ is the so called *unity tensor*. It enables to define the following *canonical isomorphism*

$$\iota : \mathbf{A} \in T_1^1(W_n) \mapsto \iota(\mathbf{A}) = L_\mathbf{A} \in L(W_n),$$
$$\mathbf{u} \in W_n \Rightarrow L_\mathbf{A}(\mathbf{u}) = \mathbf{A}\mathbf{u}. \tag{141}$$

We have

$$L_\mathbf{A} \circ L_\mathbf{B} = L_{\mathbf{AB}}, \qquad \det \mathbf{A} = \det L_\mathbf{A}, \qquad L_\mathbf{1} = I \equiv \mathrm{id}_{W_3}. \tag{142}$$

Let us denote

$$G_{af}(W_n) = W_n \times GT_1^1(W_n),$$
$$GT_1^1(W_n) := \{\mathbf{A} \in T_1^1(W_n): \det \mathbf{A} \neq 0\}. \tag{143}$$

The multiplication rule (140) defines in the set $GT_1^1(W_n)$ the structure of a group and enables to define in the set $G_{af}(W_n)$ the following algebraic structure of the semidirect product of groups $GT_1^1(W_n)$ and $(W_n, +)$ with the unit element $e_{af}$:

$$(\mathbf{a}, \mathbf{A})(\mathbf{b}, \mathbf{B}) = (\mathbf{a} + \mathbf{A}\mathbf{b}, \mathbf{A}\mathbf{B}), \qquad e_{af} = (\mathbf{o}, \mathbf{1}),$$
$$(\mathbf{a}, \mathbf{A})(\mathbf{a}, \mathbf{A})^{-1} = e_{af}, \qquad (\mathbf{a}, \mathbf{A})^{-1} = (-\mathbf{a}\mathbf{A}^{-1}, \mathbf{A}^{-1}). \tag{144}$$

We will denote by Af($W_n$) the set of all *nonsingular affine mappings* of the vector space $W_n$, that is, the bijections (i.e., mappings onto and one-one) $f: W_n \to W_n$ acting according to the rule

$$f(\mathbf{x}) = f_{(\mathbf{a}, \mathbf{A})}(\mathbf{x}) = \mathbf{A}\mathbf{x} + \mathbf{a}, \qquad (\mathbf{a}, \mathbf{A}) \in G_{af}T_1^1(W_n), \tag{145}$$

and with the group structure defined by:

$$f_{(\mathbf{a}, \mathbf{A})} \circ f_{(\mathbf{b}, \mathbf{B})} = f_{(\mathbf{a}, \mathbf{A})(\mathbf{b}, \mathbf{B})}, \qquad \mathrm{id}_{W_3} = f_{e_{af}}. \tag{146}$$

Let us denote by $(A_n, W_n, \varphi)$ the triple of geometrical objects defined by the n-dimensional real vector space $W_n$ and a set $A_n$ such that there exists a *bijection*

$$\varphi: A_n \to W_n, \tag{147}$$

The space $A_n$ is called a n-dimensional *point space* modelled on the vector space $W_n$. If $\Sigma$ is the set of all bijections of the set $A_n$ into itself, then we can define the following representation of the group $G_{af}(W_n)$ in $\Sigma$ [38]:

$$\Phi: G_{af}(W_n) \to \Sigma,$$
$$\Phi: (\mathbf{a}, \mathbf{A}) \mapsto \Phi(\mathbf{a}, \mathbf{A}) := \Phi_{(\mathbf{a}, \mathbf{A})} := \varphi^{-1} \circ f_{(\mathbf{a}, \mathbf{A})} \circ \varphi, \tag{148}$$
$$\Phi_{(\mathbf{a}, \mathbf{A})} \circ \Phi_{(\mathbf{b}, \mathbf{B})} = \Phi_{(\mathbf{a}, \mathbf{A})(\mathbf{b}, \mathbf{B})}, \qquad f_{(\mathbf{a}, \mathbf{A})} \in \mathrm{Af}(W_n).$$

Let us denote

$$l: A_n \times A_n \to W_n,$$
$$\overrightarrow{XY} \equiv l(X, Y) := \varphi(Y) - \varphi(X). \qquad (149)$$

Then

$$(l1) \quad \overrightarrow{XY} = -\overrightarrow{YX},$$
$$(l2) \quad \overrightarrow{XZ} + \overrightarrow{ZY} = \overrightarrow{XY}, \qquad (150)$$
$$(l3) \quad P, X, Y \in A_3, \quad \overrightarrow{PX} = \overrightarrow{PY} = \mathbf{x} \in W_3 \Rightarrow X = Y.$$

A *surjection* (i.e., onto) $l: A_n \times A_n \to W_n$ defined by the conditions (*l1*) - (*l3*) is called an *affine structure* in $A_n$ and the point $O = \varphi^{-1}(\mathbf{o}) \in A_n$, where $\mathbf{o} \in W_n$ is the neutral element (*zero vector*) of $W_n$, is called an *origin* of $A_n$. The set $A_n$ of eq.(149) endowed with an affine structure is called the *affine point space* modelled on $W_n$ and the bijection $\varphi$ is called then an *affine structural bijection* (of the affine point space $A_n$). Notice that if $l$ is an affine structure in $A_n$ and $O \in A_n$ is an arbitrary established point, then the mapping

$$\varphi_O: A_n \to W_n \qquad (151)$$

defined by

$$X \in A_n \mapsto \varphi_O(X) = \overrightarrow{OX} \in W_n \qquad (152)$$

is bijective, fulfils the condition

$$l(X, Y) = \varphi_O(Y) - \varphi_O(X), \qquad (153)$$

and the condition

$$\varphi_O^{-1}(\mathbf{o}) = O. \qquad (154)$$

Thus, we can identify, up to the choice of a point $O \in A_n$, a bijection $\varphi$ fulfilling the conditions (149) and (150) and the mapping $\varphi_O$ defined by eqs.(151) and (152). A mapping $\varphi_P$, $P \in A_n$, is called the *localization* of the affine structure at the point P (or the *radius-vector* fastened at this point).

The mapping $\psi$ defined by:

$$\psi: (W_n, +) \to \Sigma, \quad \psi(\mathbf{a}) = \psi_\mathbf{a},$$
$$\psi_\mathbf{a} = \Phi(\mathbf{a}, \mathbf{1}): A_n \to A_n, \quad \mathbf{a} \in W_n, \qquad (155)$$

defines an action of the abelian group $(W_n, +)$ in $A_n$, that is

$$\psi_\mathbf{a} \circ \psi_\mathbf{b} = \psi_{\mathbf{a+b}} = \psi_\mathbf{b} \circ \psi_\mathbf{a} \qquad (156)$$

The group $(W_n, +)$ is called the group of *translations* and acts not only *transitively* (i.e., for each pair of points $(X, Y) \in A_n \times A_n$ there exists a vector $\mathbf{a} \in W_n$ such that $\psi_\mathbf{a}(X) = Y$) but also *freely* in the sense that isotropy groups of all points are trivial. [42]

Let Af($A_n$) denotes the set of all mappings $f : A_n \to A_n$ preserving the affine structure $l$. These mappings are called the *affine mappings* and induces the linear mappings $L(f) \in L(W_n)$ such that, for each pair of points $(X, Y) \in A_n \times A_n$, the following condition is fulfilled:

$$\overrightarrow{f(X)f(Y)} = L(f)(\overrightarrow{XY}). \tag{157}$$

Let us denote

$$\begin{aligned} \text{Af}(A_n)_O &= \{f \in \text{Af}(A_n): f(O) = O\}, \\ T(A_n) &= \{f \in \text{Af}(A_n): f = \psi_{\mathbf{a}}, \text{ for certain } \mathbf{a} \in W_n\}. \end{aligned} \tag{158}$$

Then all mappings $f \in \text{Af}(A_n)$ are of the form

$$f = t \circ f_O, \quad t \in T(A_n), \quad f \in \text{Af}(A_n)_O, \quad O \in A_n, \tag{159}$$

that is, for each point $X \in A_n$, we have:

$$\overrightarrow{Of(X)} = L(f)(\overrightarrow{OX}) + \overrightarrow{Of(O)}. \tag{160}$$

Thus, if $O \in A_n$ is arbitrary established, then the mappings $f \in \text{Af}(A_n)$ act according to the rule:

$$\overrightarrow{Of(X)} = f_{(\mathbf{a},\mathbf{A})}(\overrightarrow{OX}), \quad f_{(\mathbf{a},\mathbf{A})} \in \text{Af}(W_n), \tag{161}$$

for each $X \in A_n$, where eq.(145) was taken into account. We will designate by $GL(W_n) \subset L(W_n)$ the set of all linear bijections in $W_n$. Because $GL(W_n) = \iota(GT_1^1(W_n))$ (see eqs.(141) - (143)), we can define in this set the algebraic structure of a group induced from $GT_1^1(W_n)$.

Let us consider the pair $e(P_0) = \{P_0, e\}$, $P_0 \in A_n$ - a distinguished point, $e = \{\mathbf{e}_i\}_{1 \leq i \leq n}$ - a base of $W_n$. The such defined pair $e(P_0)$ is called an *affine frame* in $A_n$ localized in $P_0$ and associated with the (local) base $e$. If $\varphi_{P_0}$ is the localization of the affine structure at the point $P_0$ (see eqs.(151) - (154)), and $\mathbf{r}: A_n \to W_n$ is the radius-vector fastened at this point, then the formula

$$\mathbf{r}(P) \equiv \varphi_{P_0}(P) = x^i(P)\mathbf{e}_i, \quad P \in A_n, \tag{162}$$

defines the following *affine coordinate system* on the point space $A_n$:

$$\begin{aligned} x &: A_n \to \mathbb{R}^n, \\ x(P) &= (x^i(P); i \to 1,\ldots,n), \quad x(P_0) = 0 \equiv (0,\ldots,0) \in \mathbb{R}^n, \end{aligned} \tag{163}$$

associated with the affine frame $e(P_0)$. If $e'(P_0') = (P_0', e')$, $e' = (\mathbf{e}_i')_{1 \leq i \leq n}$, $\overrightarrow{P_0 P_0'} = \mathbf{a} = a^i \mathbf{e}_i \in W_n$, is a second affine frame, then the mapping $f \in \text{Af}(A_n)$ such that $e'(P_0') = f(e(P_0))$ is uniquely de-fined and if $y \equiv x' : A_n \to \mathbb{R}^n$ is a new affine coordinate system, then the coordinate description of this mapping is given by the following bijective affine mapping acting in the arithmetic space $\mathbb{R}^n$:

$$y^i = A_j^{\,i} x^j + a^i, \qquad \mathbf{A} = \left\| A_j^{\,i} \,{}_{j \to 1,2,\ldots,n}^{i \downarrow 1,2,\ldots,n} \right\| \in \mathrm{GL}(n) \ . \tag{164}$$

If the vector space $W_n$ is oriented, then the set of matrices

$$\mathrm{GL}^+(n) = \{ \mathbf{A} \in \mathrm{GL}(n) : \det \mathbf{A} > 0 \} \tag{165}$$

defines the orientation preserving affine mappings.

**Appendix B - Differential geometry of affine spaces**

Let us denote by $\Psi_n$ the set of all affine coordinate systems on the affine point space $A_n$ modelled on the vector space $W_n$ (Appendix A). The *topology* $\tau_n$ in $A_n$ is defined as a weaker topology in which the affine coordinate systems on $A_n$ are continuous mappings. It is the family of the following sets:

$$\tau_n = \left\{ x^{-1}(U) : U \in \mathrm{top}\,\mathbb{R}^n, \ x \in \Psi_n \right\}. \tag{166}$$

Consequently, the set $\Psi_n$ can be considered as an *atlas* that defines the *differential structure* $\Omega_n$ of the affine space $A_n$ consisting of all real functions $f$ on $A_n$ such that [39]

$$\Omega_n = \left\{ f : A_n \to \mathbb{R}, \quad f \circ x^{-1} \in C^\infty(\mathbb{R}^n) \ \text{for each} \ x \in \Psi_n \right\}. \tag{167}$$

The affine space endowed with this differential structure is a *differential manifold*.

Let's notice that if the affine point space $A_n$ is considered as a differential manifold and $T_P A_n$ is the vector space tangent to $A_n$ at the point $P \in A_n$, then the localization $\varphi_P$ of $A_n$ enables to identify the space $T_P A_n$ and $W_n$ in this sense, that (see Appendix A):

$$T_P A_n \simeq \varphi_P(A_n) = \left\{ \overrightarrow{PQ} \in W_n, \ Q \in A_n \right\} \equiv (P, W_n) \simeq W_n \ . \tag{168}$$

The affine frame $e(P) = \{P, e\}$, $e = \{\mathbf{e}_i\}_{1 \le i \le n}$, localized at the point P is then a base of $T_P A_n$ and can be identified with the system $E(P) = \{P, P_1, \ldots, P_n\}$ of points uniquely defined by the condition

$$\mathbf{e}_i = \overrightarrow{PP_i} \in \varphi_P(A_n), \qquad i = 1,\ldots,n \ . \tag{169}$$

Let $U \subset A_n$ be an open subset and let $W(U)$ denote the module of all smooth vector fields on $U$ tangent to $U$ (see [45] - Appendix). These fields can be identified, according to eq.(168), with the mappings $\mathbf{v} : U \to W_n$. However, it is frequently more convenient to consider these fields as the first-order linear differential operators defined at each point $P \in U$ as the operators $\partial_{\mathbf{v}_P}$ of differenti-ating of smooth functions $f \in C^{\infty}(U)$ in the direction $\mathbf{v}_P \equiv \mathbf{v}(P) \in W_n$ (see, for more precise reasoning, [45] - Appendix).

Let $ST_2(W_n) \subset T_2(W_n) = W_n^* \otimes W_n^*$ denotes the linear space of all symmetric 2-covariant tensors. A distinguished tensor $\mathbf{g} \in ST(W_n)$ is called the *fundamental tensor* of $W_n$ and defines, independently from the choice of the base $e = \{\mathbf{e}_i\}_{1 \leq i \leq n}$ of the vector space $W_n$, the R-bilinear symmetric mapping $(\cdot,\cdot)_g : W_n \times W_n \to R$ acting according to the rule:

$$\mathbf{u} \cdot \mathbf{v} \equiv \mathbf{ugv} \equiv (\mathbf{u}, \mathbf{v})_g = g_{ij} u^i v^j,$$
$$\mathbf{g} = g_{ij} \mathbf{e}^i \otimes \mathbf{e}^j, \quad \mathbf{u} = u^i \mathbf{e}_i, \quad \mathbf{v} = v^j \mathbf{e}_j, \quad (170)$$

where $e^* = (\mathbf{e}^i)_{1 \leq i \leq n}$ is the base of the covector space $W_n^*$ dual to the base $e$ of $W_n$. The space $W_{n,g} = (W_n, \mathbf{g})$ is called *orthogonal*. The orthogonal space is called *pseudo-Euclidean* (or *nondegenerate*) if the mapping

$$G: W_n \to W_n^*,$$
$$v = G(\mathbf{v}) := (\mathbf{v}, \cdot)_g, \quad (171)$$

is *injective*, that is for any $0 \neq \mathbf{v} \in W_n$ there exists $\mathbf{u} \in W_n$ such that $(\mathbf{v}, \mathbf{u})_g \neq 0$. If the mapping $G$ is not injective, then the space $W_{n,g}$ is called *degenerate*. In pseudo-Euclidean spaces the mapping $G$ defines a *canonical isomorphism*.[27] Particularly, if $e = (\mathbf{e}_i)_{1 \leq i \leq n}$ is a base of $W_n$ and $e^* = (\mathbf{e}^i)_{1 \leq i \leq n}$ is the base of $W_n^*$ dual to $e$, then this dual base can identified with the base $e_G = (\mathbf{e}^i)_{1 \leq i \leq n}$ of $W_n$ defined by the condition

$$\mathbf{e}^i = G(\mathbf{e}^i), \quad i = 1,...,n. \quad (172)$$

If the affine point space $A_n$ is considered as a differential manifold whose tangent spaces are identified with an orthogonal vector space $W_{n,g} = (W_n, \mathbf{g})$, then this space is called the *orthogonal point space modelled on* $W_{n,g}$ and can be identified with the (semi-)Riemannian space denoted as $A_{n,g}$. In this paper are considered three-dimensional orthogonal point spaces Minkowski or Euclidean type. These spaces can be defined by the condition of the existence such $\mathbf{g}$-orthonormal vector base $e_\varepsilon = (\mathbf{e}_i)_{1 \leq i \leq 3}$ of the space $W_{3,g} = (W_3, \mathbf{g})$ that if

$$g_{ij} = (\mathbf{e}_i, \mathbf{e}_j)_g, \quad (173)$$

then

$$\mathbf{G}_\varepsilon = \left\| g_{ij}; \begin{smallmatrix} i \downarrow 1,2,3 \\ j \to 1,2,3 \end{smallmatrix} \right\| = \text{diag}(1,1,\varepsilon), \quad (174)$$

where $\varepsilon = 1$ for the Euclidean space and $\varepsilon = -1$ for the Minkowski space. In the case of orthogonal Minkowski spaces we are dealing with three kinds of vectors $0 \neq \mathbf{v} \in W_{3,g}$:

$$\begin{aligned}(\mathbf{v}, \mathbf{v})_g &> 0 \quad - \quad \text{space-like vector,} \\ (\mathbf{v}, \mathbf{v})_g &< 0 \quad - \quad \text{timelike vector,} \\ (\mathbf{v}, \mathbf{v})_g &= 0 \quad - \quad \text{isotropic (or null) vector.}\end{aligned} \quad (175)$$

The null vectors form a cone in $W_3$ called the *null cone* („*light cone*" in physics).

Let us consider a smooth *immersion* $\kappa: M \to A_3$ of a two-dimensional smooth manifold $M$ into the three-dimensional affine point space $A_3$ modelled on an orthogonal vector space $W_{3,g}$ (Euclidean or Minkowski type). It means that the induced *tangent mapping* $\kappa_*: W(M) \to W(A_3)$, where W($M$) is the module of smooth vector fields tangent to $M$ (see e.g. [45] - Appendix) and $W(A_3)$ is the module of smooth vector fields tangent to $A_3$ with values in the considered orthogonal vector space, is *injective*. An immersion is not necessarily injective. An injective immersion is an *embedding*. The set $\kappa(M)$ with the differentiable structure induced by the embedding $\kappa$ is a *manifold*. The differentiable structure on $\kappa(M)$ induced by $\kappa$ is the set of charts $\{(K(U), u \circ K^{-1})\}$ where $\{(U, u)\}$ is an atlas on $M$ and $K^{-1}: \kappa(M) \to M$ is a bijection such that $K^{-1} \circ \kappa = \mathrm{id}_M$; the mapping $K$ differs from $\kappa$ in that it is surjective. [4]

The *manifold structure* induced by $\kappa$ on $\kappa(M)$ may not be equivalent to a *submanifold structure* on $\kappa(M)$ treated as a space endowed with the topology induced from $A_3$. If $\kappa(M)$ has a submanifold structure equivalent to the manifold structure induced by the embedding, then the embedding is said to be *regular* [4, 17]. If $\kappa$ is a regular embedding, then $\kappa(M)$ is a *submanifold* of $A_3$ called the *regular surface* [17].

**Appendix C - Differential operators** [4, 39, 50]

Let $M_a = (M, \mathbf{a})$ be a m-dimensional (semi-)Riemannian space. The (semi-)metric $\mathbf{a}$ induces many operations called „raising" and „lowering" indices. Namely, a non-singular metric tensor induces at each point $P \in M$ the non-singular mapping $G$ of $T_P M$ onto $T_P^* M$ acting according to the rule

$$\begin{aligned}\mathbf{u}_P \in T_P M &\mapsto G(\mathbf{u}_P) \in T_P^* M, \\ G(\mathbf{u}_P)(\mathbf{v}_P) &= (\mathbf{u}_P, \mathbf{v}_P)_a, \quad \mathbf{v}_P \in T_P M.\end{aligned} \quad (176)$$

We let $G^*$ denote the inverse map of $T_P^* M$ onto $T_P M$ acting according to the rule

$$u_P \in T_P^*M \mapsto G^*(u_P) \in T_PM,$$
$$\left(G^*(u_P), \mathbf{v}_P\right)_a = u_P(\mathbf{v}_P), \quad \mathbf{v}_P \in T_PM.$$
(177)

If $\varphi \in C^\infty(M)$, then we can define the *gradient field* of $\varphi$ by the rule

$$\operatorname{grad}_a \varphi = G^*(d\varphi),$$
$$d\varphi(\mathbf{u}) = \partial_\mathbf{u}\varphi, \quad \mathbf{u} \in W(M),$$
(178)

and the *Laplacian operator* by

$$\Delta_a f = \operatorname{div}_a(\operatorname{grad}_a f),$$
$$\operatorname{div}_a \mathbf{u} = |a|^{-1/2} \partial_k \left(|a|^{1/2} u^k\right), \quad \mathbf{u} = u^k \partial_k \in W(M).$$
(179)

It follows that in local coordinates $x = (x^1,...,x^m)$ on $M$

$$\Delta_a f = |a|^{-1/2} \partial_i \left(|a|^{1/2} a^{ij} \partial_j f\right) = a^{ij} \partial_i \partial_j f - \Gamma^k \partial_k f,$$
(180)

where

$$\Gamma^k = a^{ij} \Gamma^k_{ij} = -|a|^{-1/2} \frac{\partial}{\partial x^i}\left(|a|^{1/2} a^{ik}\right),$$
(181)

$\Gamma^k_{ij}$ are Christoffel symbols, and $a = \det \mathbf{a} \doteq \det(a_{ij})$.

If $M$ and $N$ are smooth differential manifolds (m-dimensional and n-dimensional, respectively), and $f : M \to N$ is a smooth mapping, then the *differential* of $f$ at the point $P \in M$ is a R-linear mapping defined by:

$$df_P \equiv d_P f : T_PM \to T_{f(P)}N,$$
$$d_P f(\mathbf{v}_P)(\alpha) = \mathbf{v}_P(f^*\alpha), \quad \mathbf{v}_P \in T_PM,$$
$$f^*\alpha \equiv \alpha \circ f \in C^\infty(N), \quad \alpha \in C^\infty(M).$$
(182)

If additionally $f$ is a surjective diffeomorphism, then we can define the *tangent mapping* (also called the differential) *induced* by $f$ as a R-linear mapping $f_*$ acting according to the following rule:

$$f_* : W(M) \to W(N),$$
$$f_*\mathbf{v} = df \circ \mathbf{v} \circ f^{-1}, \quad \mathbf{v} \in W(M),$$
(183)

where

$$df : P \in M \mapsto df(P) := df\big|T_PM = d_P f.$$
(184)

The *reciprocal image* (*pull back*) $f^*\omega$ of the form $\omega$ is defined by

$$f^*\omega(\mathbf{u}_1,...,\mathbf{u}_k) = \omega(f_*\mathbf{u}_1,...,f_*\mathbf{u}_k),$$
$$\mathbf{u}_i \in W(M), \quad i = 1,...,k \leq m = \dim M, \quad k \leq n = \dim N.$$
(185)

The reciprocal image $f^*\omega$ is also called the *form induced* by $f$ from $\omega$.

In the paper are considered, because of physical interpretations of geometrical objects, *dimensional coordinate systems* $(U, x)$ on the differential manifold $M$ with coordinates $x = (x^i)$ of the dimension of length, i.e., $[x^i]$ = cm. If the corresponding natural base vectors $\mathbf{e}_i \in W(U)$, $1 \leq i \leq m$, tangent to the coordinate lines are identified with differentiations in the directions of these vectors, i.e.,

$$\mathbf{e}_i(f) \equiv \partial_{\mathbf{e}_i} f = \partial f / \partial x^i, \quad f \in C^\infty(M) \tag{186}$$

(see [45] - Appendix), then would be $[\mathbf{e}_i]$ = cm$^{-1}$ and $[\mathbf{e}^i] = [\mathrm{d}x^i]$ = cm. In this case, we have:

$$\mathbf{v} = v^i \mathbf{e}_i \in W(U), \quad [\mathbf{v}] = \mathrm{cm}^{-1} \Rightarrow [v^i] = [1], \tag{187}$$

or, for example,

$$\mathbf{v} = v^i \mathbf{e}_i \in W(U), \quad [v^i] = \mathrm{cm} \Rightarrow [\mathbf{v}] = [1]. \tag{188}$$

**Acknowledgements**

This paper contains results obtained within the framework of the research project N N501 049540 financed from Scientific Research Support Fund in 2011-2014. The author is greatly indebted to the Polish Ministry of Science and Higher Education for this financial support.